\documentclass[fleqn,usenatbib]{mnras}

\usepackage[T1]{fontenc}
\usepackage{ae,aecompl}

\usepackage{times}

\usepackage{graphicx}	
\usepackage{amsmath}	
\usepackage{amssymb}	

\usepackage{subcaption}
\usepackage{url}

\newcommand{\vect}[1]{\boldsymbol{#1}}
\newcommand{\diff}{\text{d}}
\newcommand{\imag}{\text{i}}
\newcommand{\e}{\text{e}}

\date{}
\pubyear{2020}
\title[Tidal oscillations in hot WDs]{Tidally excited oscillations in hot white dwarfs}

\author[H. Yu, J. Fuller and K. B. Burdge]{Hang Yu$^{1,2}$\thanks{E-mail: hangyu@caltech.edu}, Jim Fuller$^{1,2}$ and Kevin B. Burdge$^{2}$ \\
$^{1}$TAPIR, Walter Burke Institute for Theoretical Physics, Mailcode 350-17 California Institute of Technology, Pasadena, CA 91125, USA\\
$^{2}$Division of Physics, Mathematics and Astronomy, California Institute of Technology, Pasadena, CA, USA}

\begin{document}

\defcitealias{Weinberg:12}{WAQB12}
\defcitealias{Fuller:12a}{FL12}
\defcitealias{Burkart:13}{BQAW13}
\defcitealias{Yu:20}{YWF20}

\label{firstpage}
\pagerange{\pageref{firstpage}--\pageref{lastpage}}
\maketitle

\begin{abstract}
We study the flux variation in helium white dwarfs (WDs) induced by dynamical tides for a variety of WD models with effective temperatures ranging from $T=10\,{\rm kK}$ to $T=26\,{\rm kK}$. 
At linear order, we find the dynamical tide can significantly perturb the observed flux in hot WDs. If the temperature $T\gtrsim 14\,{\rm kK}$, then the dynamical tide may induce a fractional change in the observed flux by $>1\%$ when the orbital period is $P_{\rm orb}\simeq 20-60\,{\rm min}$. 
The ratio between the flux modulation due to the dynamical tide and that due to the equilibrium tide (i.e., ellipsoidal variability) increases as the WD's radius decreases, and it could exceed $\mathcal{O}(10)$ if the WD has a radius $R\lesssim 0.03 R_\odot$. Unlike the ellipsoidal variability which is in phase with the orbital motion, the pulsation caused by the dynamical tide may have a substantial phase shift. 
A cold WD with $T\simeq 10\,{\rm kK}$, on the other hand, is unlikely to show observable pulsations due to the dynamical tide.
At shorter orbital periods, the dynamical tide may break and become highly nonlinear. We approximate this strongly nonlinear regime by treating the waves as one-way traveling waves and find the flux variation is typically reduced to $0.1\%-1\%$ and the excess phase is likely to be $90^\circ$ (though with large uncertainty). Even in the traveling wave limit, the flux perturbation due to dynamical tide could still exceed the ellipsoidal variability for compact WDs with $R\lesssim 0.02 R_\odot$.
We further estimate the nonlinear flux perturbations oscillating at four times the orbital frequency dominated by a self-coupled parent g-mode driving low-order daughter p-modes. 
The nonlinear flux variation could be nearly $50\%$ of the linear variation for very hot WD models with $T\gtrsim26\,{\rm kK}$ and $1\%$ linear flux variation. 
We thus predict that both the linear and nonlinear flux variations due to dynamical tides are likely to have significant observational signatures for hot WDs in compact binaries. 
\end{abstract}

\begin{keywords}
(stars:) binaries (including multiple): close -- stars: oscillations (including pulsations) -- white dwarfs -- asteroseismology
\end{keywords}

\section{INTRODUCTION}
\label{sec:intro}

Binary white dwarfs (WDs) are the raw ingredients needed to produce a large menu of astrophysical phenomena. AM CVn systems \citep{Nelemans:01}, R Cor Bor Stars \citep{Clayton:12}, rapidly rotating magnetic white dwarfs \citep{Ferrario:15}, accretion induced collapse, and various classes of type Ia supernovae very likely originate from mass transferring and/or merging WDs \citep{Webbink:84,Iben:84,Toonen:12,Shen:18,Polin:19,Polin:20}. Short-period WD systems are also the primary source of gravitational waves (GWs) that will be detectable by the  Laser Interferometer Space Antenna (LISA, \citealt{Amaro-Seoane:17}), and other space-based GW observatories (TianQin, \citealt{Luo:16}; and TianGO, \citealt{Kuns:19})

By the time LISA flies, however, ground-based surveys will have already detected and characterized dozens (if not more) of short-period ($P_{\rm orb} < 1 \, {\rm hr}$) WD binaries. Over the last decade, the ELM survey \citep{Brown2020b} has detected roughly a dozen such short-period WD binaries, while the Zwicky Transient Facility (ZTF, \citealt{Bellm2019,Masci2019,Graham2019,Dekany:20}) has uncovered roughly a dozen more in the last two years alone \citep{Burdge:20}. Time-domain surveys like ZTF are especially powerful for finding short-period systems, where photometric variability due to eclipses and/or ellipsoidal modulation is most likely to be detectable. The Vera Rubin Observatory (VRO, \citealt{Ivezic2019}) time-domain survey will extend significantly deeper than ZTF and will further expand the sample of short-period WDs that can be characterized by electromagnetic observations. 

An important facet of these systems is the tidal interaction that strengthens as the orbital separation of the WDs decreases due to GW orbital decay. The dominant form of tidal energy dissipation is expected to be tidally excited gravity (i.e., buoyancy) waves within the WDs, which are dissipated via radiative diffusion and non-linear processes \citep{Fuller:11,Fuller:12a,Fuller:12b,Burkart:13,Fuller:13,Fuller:14,Yu:20}. Consequently, the WDs are expected to be nearly tidally synchronized by the time they begin mass transfer at very short orbital periods ($P_{\rm orb} \! \lesssim 10 \, {\rm min}$), and tidal heating could make the WDs substantially hotter and brighter. Crucially, tidal energy dissipation is expected to cause the orbit to decay several percent faster than GW emission alone, an effect that will likely be detectable via GW observations or eclipse timing \citep{Piro:19,Yu:20}. 

However, nearly all the above work has neglected to address an important observable feature of short-period WD binaries, namely that of photometric modulation due to tidally excited oscillation modes within the WDs. Here, we show that tidally excited oscillations are expected to be detectable in many short-period ($P_{\rm orb} \lesssim 60 \, {\rm min})$ binaries. These oscillations offer a new route to quantify the importance of tidal effects, and they may be much easier to detect than the tidal synchronization or heating that results from the dissipation of these oscillations. Furthermore, we show that these oscillations will be most pronounced in hot WDs ($T \gtrsim 14,000 \, {\rm K}$) in which the tidally excited gravity modes can propagate close to the surface. The resulting tidal dynamics can also differ greatly, as the radiative and non-linear gravity mode damping processes are strongly altered in hot WDs compared to the cooler WD models considered in prior work.

\section{WD MODELS}
\label{sec:models}
\begin{table}
\begin{center}
\caption{\label{tab:bgConfig}The mass $M$, effective temperature $T $, radius $R$, moment of inertia $I_{\rm WD}$, natural energy $E_0{\equiv} GM^2/R$, and natural frequency $\omega_0{\equiv}\sqrt{GM/R^3}$ of our WD models. We have used the cgs units for dimensional quantities.}
\begin{tabular}{cccccc}
$M/M_\odot$ & $T $  &  $R/R_\odot$ &      
$I_{\rm WD}/MR^2$ & $E_0$  &  $\omega_0$   \\
\hline
$0.4 $ & $26\,{\rm k}$ & $0.023$ & $0.18$ & $2.6\times10^{49}$ & $0.11$ \\
$0.25$ & $18\,{\rm k}$ & $0.036$ & $0.16$ & $6.7\times10^{48}$ & $0.047$\\
$0.25$ & $14\,{\rm k}$ & $0.027$ & $0.20$ & $8.8\times10^{48}$ & $0.070$\\
$0.25$ & $10\,{\rm k}$ & $0.023$ & $0.24$ & $1.0\times10^{49}$ & $0.091$
\end{tabular}
\end{center}
\end{table}

In this study we consider four He WD models. One has a mass of $M=0.4\,M_\odot$ and effective temperature $T =26\,{\rm kK}$. The other three all have the same mass $M=0.25\,M_\odot$ and respectively have $T =18\,{\rm kK},\ 14\,{\rm kK},\ 10\,{\rm kK}$. The bulk properties of these models are summarized in Table~\ref{tab:bgConfig}. For future convenience, we will refer to these models according to their effective temperature as T26, T18, T14, and T10, respectively. 

These background models are constructed using \texttt{MESA} (version 12115;~\citealt{Paxton:11, Paxton:13, Paxton:15, Paxton:18, Paxton:19}). Specifically, we start from a zero-age main-sequence star with $\simeq 2\,M_{\odot}$ and evolve it until its helium core reaches the desired WD mass. We then strip all but approximately $10^{-4}$ in mass of the hydrogen and cool the model to the desired effective temperature. 
In the cooling process we do not include element diffusion in our models.

\begin{figure*}
\subfloat[$M=0.4\,M_\odot$, $T=26\,{\rm kK}$]{
	\begin{minipage}[c][0.915\width]{0.48\textwidth}
	   \centering
	   \includegraphics[width=\textwidth]{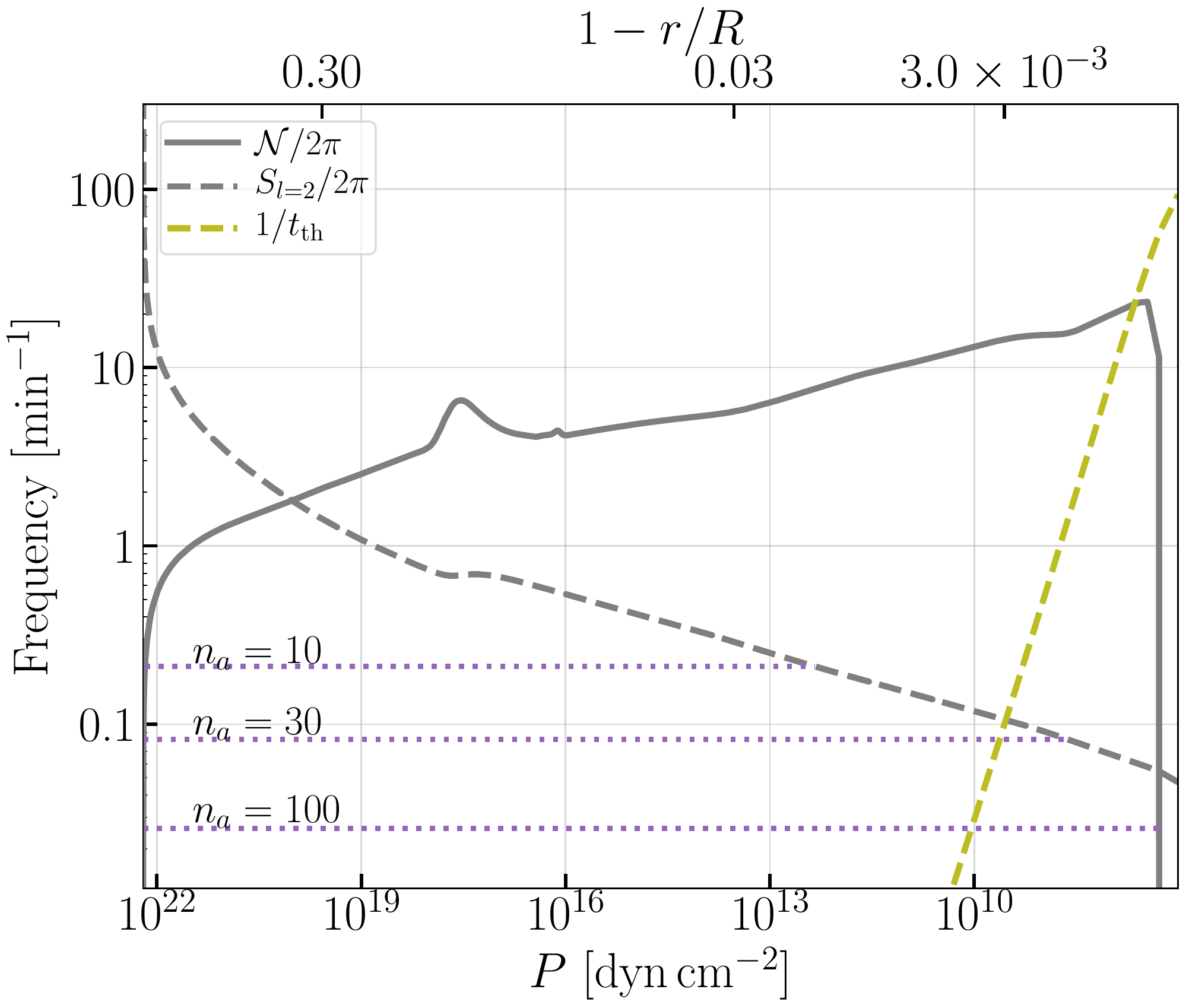}
	\end{minipage}}
 \hfill 	
  \subfloat[$M=0.25\,M_\odot$, $T=18\,{\rm kK}$]{
	\begin{minipage}[c][0.915\width]{0.48\textwidth}
	   \centering
	   \includegraphics[width=\textwidth]{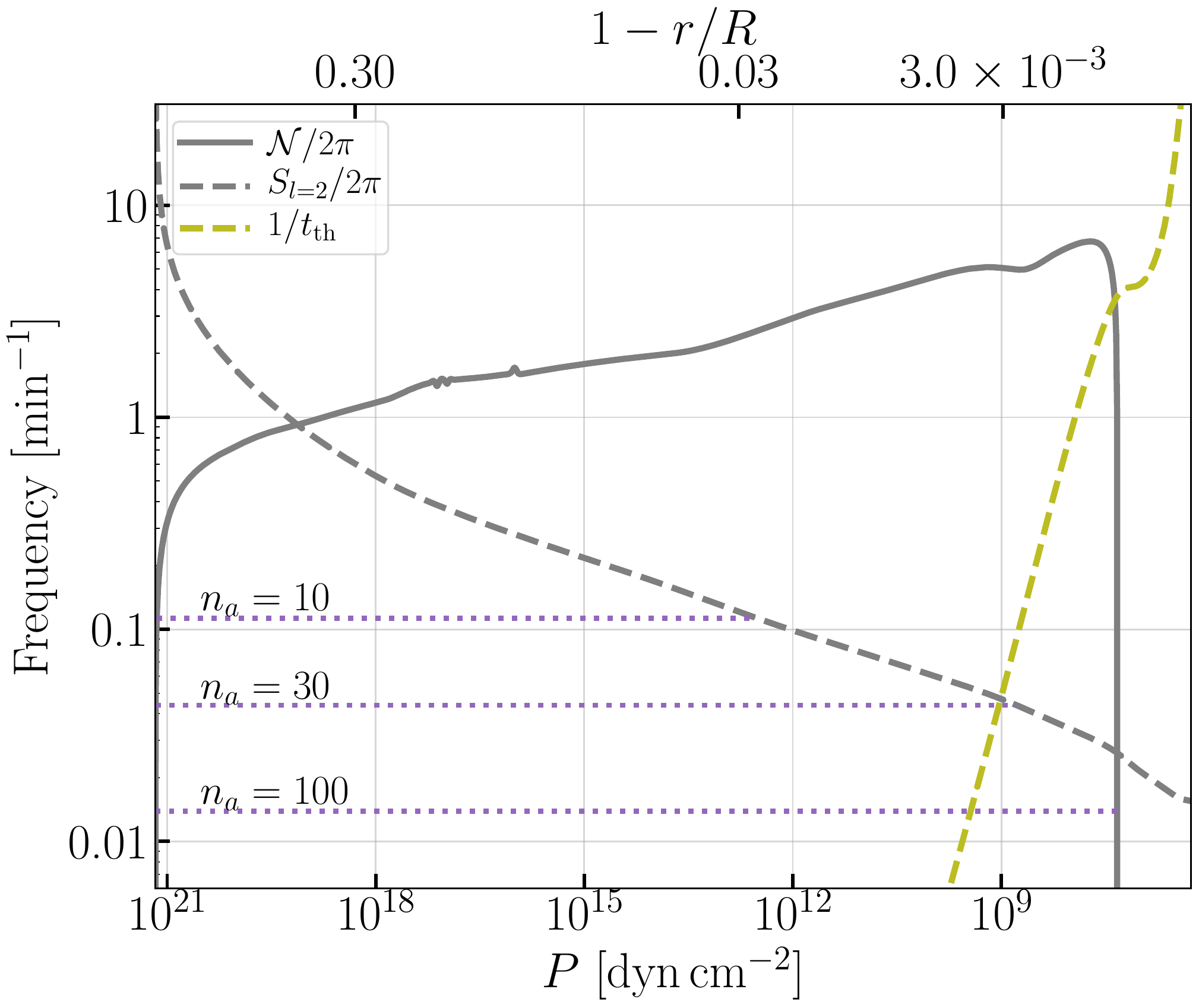}
	\end{minipage}}
 \hfill	
  \subfloat[$M=0.25\,M_\odot$, $T=14\,{\rm kK}$]{
	\begin{minipage}[c][0.915\width]{0.48\textwidth}
	   \centering
	   \includegraphics[width=\textwidth]{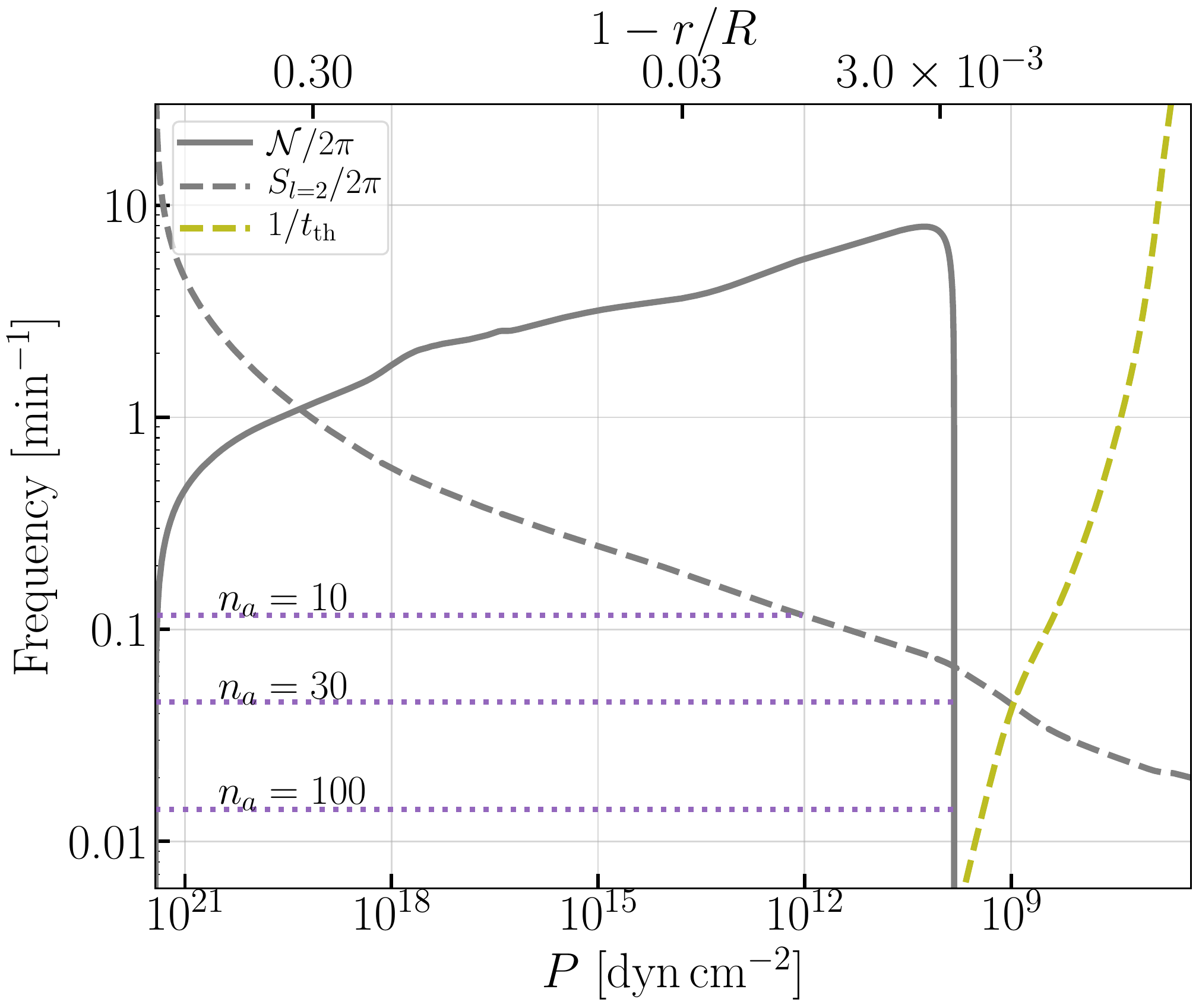}
	\end{minipage}}
\hfill	
  \subfloat[$M=0.25\,M_\odot$, $T=10\,{\rm kK}$]{
	\begin{minipage}[c][0.915\width]{0.48\textwidth}
	   \centering
	   \includegraphics[width=\textwidth]{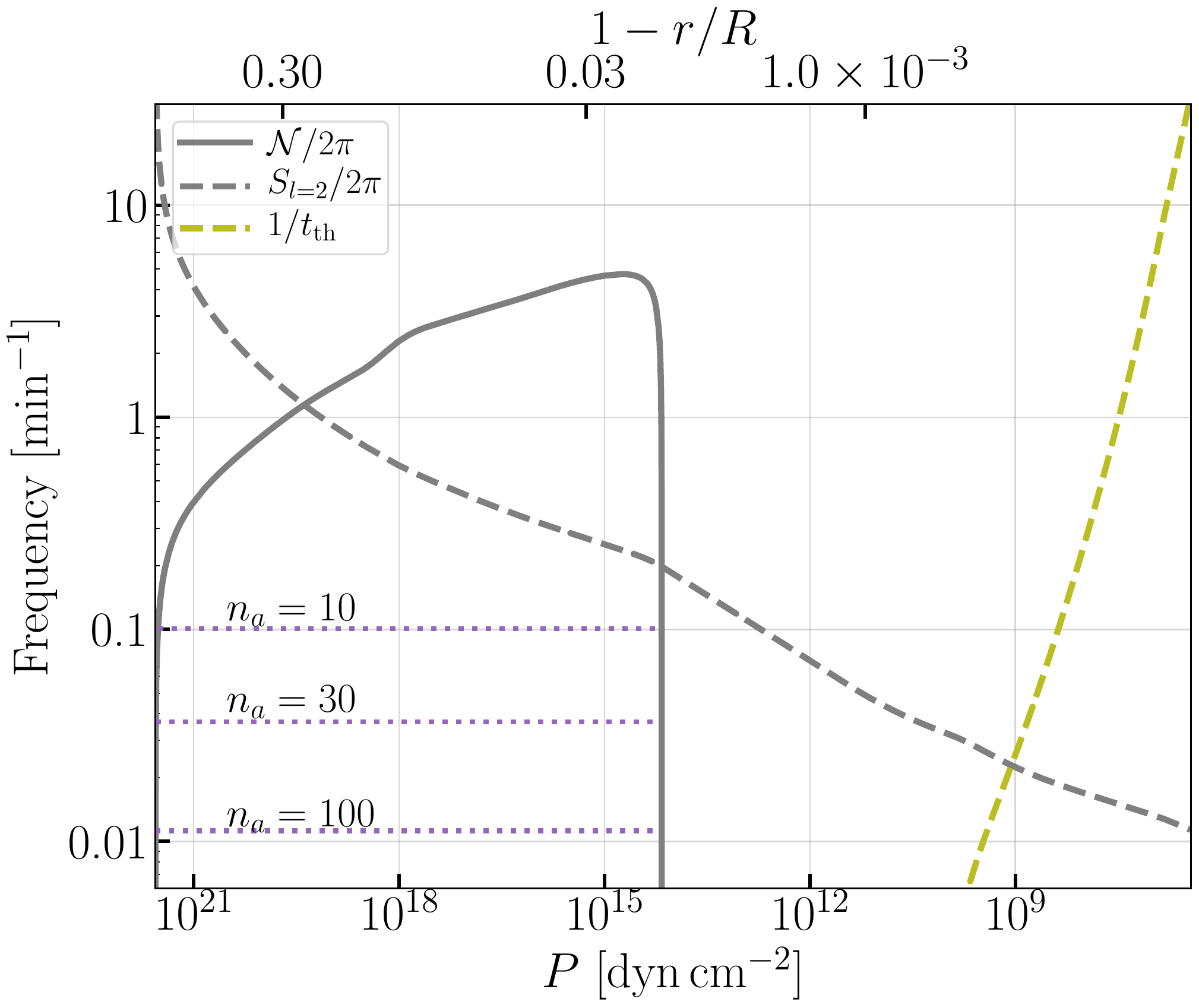}
	\end{minipage}}
\caption{Propagation diagrams of the four WD models. We use solid-grey and dashed-grey traces to represent the Brunt-V{\"a}is{\"a}l{\"a} frequency and the Lamb frequency, respectively. As comparisons, the dotted-purple lines show eigenfrequencies of the $n_a=10,\,30,\,\text{and}\,100$ g-modes (with $l_a=2$ and the span of each trace corresponds to the propagation cavity of the corresponding mode. The dashed-olive trace shows the inverse of the local thermal timescale. Note we have inverted the x-axes so that the WD surface is on the right). 
}
\label{fig:combo_prop}
\end{figure*}

Figure~\ref{fig:combo_prop} shows the propagation diagrams of the models. Here we use solid-grey traces to represent the Brunt-V{\"a}is{\"a}l{\"a} frequency (i.e. buoyancy frequency)
\begin{equation}
    \mathcal{N}^2 = g^2\left(\frac{1}{c_{\rm e}^2} - \frac{1}{c_{\rm s}^2}\right),
\end{equation}
where $c_{\rm e}^2 = \diff P/\diff \rho$ is the equilibrium sound speed squared and $c_{\rm s}^2=\Gamma_{1}P/\rho$ is the adiabatic sound speed square with $\Gamma_1$  the adiabatic index. All other quantities have their usual meaning. 
The dashed-grey traces represent the Lamb frequency 
\begin{equation}
    S_{l}^2=\frac{l(l+1)c_s^2}{r^2}, 
\end{equation}
and in the plots we have set the polar degree $l=2$ as the tidal interaction is dominated by terms at the quadruple ($l=2$) order. 

Here we are particularly interested in gravity modes, or g-modes, inside the WD as they are mostly responsible for the dynamical tides. A g-mode can propagate if its eigenfrequency $\omega_a$ satisfies $\omega_a<\mathcal{N}$ and $\omega_a<S_l$. Under the Wentzel–Kramers–Brillouin (WKB) approximation, the radial wavenumber of a g-mode can be written as a function of the eigenfrequency $\omega_a$ as \citep{Christensen:98}
\begin{equation}
    k_r^2(\omega_a) = \frac{\omega_a^2}{c_{\rm s}^2}\left(\frac{S_l^2}{\omega_a^2}-1\right)\left(\frac{\mathcal{N}^2}{\omega_a^2}-1\right). 
    \label{eq:wkb_kr_sq}
\end{equation}

One interesting feature to notice is that the radiative zone (where $\mathcal{N}>0$) expands towards the surface as the temperature increases. As a result, g-modes in the hot T26 and T18 models can propagate very close to the stellar surface and they have different outer boundaries determined by the Lamb frequency. In contrast, the g-modes in the cold T10 model (similar to the model studied in \citealt{Yu:20}; hereafter, \citetalias{Yu:20}) can only propagate to a pressure of $P\simeq 10^{14}\,{\rm dyn\,cm^{-2}}$ and they all have essentially the same outer boundary set by the buoyancy frequency as $\omega_a\ll S_l$ for almost all the g-modes in such a cold WD. Furthermore, in the hot models, g-modes can reach to regions where the local thermal timescale $t_{\rm th}$ is small and the damping due to radiative diffusion is significant, where 
\begin{equation}
    t_{\rm th} = \frac{P C_{P} T}{gF},
\end{equation}
with $C_P$ the heat capacity at constant pressure and $F$ the radiative flux. As we shall see shortly in Section~\ref{sec:tide_eqs} and in Figure~\ref{fig:combo_lin_par}, the difference in the g-mode propagation cavities between hot and cold WD models will lead to distinct characteristics of g-modes in these models.


\section{Formalism}
\label{sec:formalism}

\subsection{Tidal interaction}
\label{sec:tide_eqs}
Our treatment of the tidal interaction follows closely \citetalias{Yu:20}  and references therein. Specifically, we consider the problem in a coordinate system whose origin is at the centre of the primary star with mass $ M$ and corotates with it. We assume that the orbit is circular and that the spin angular momentum of the primary is aligned with the orbital angular momentum. The linear tidal response of a fluid element inside the primary to the gravitational field of the secondary ($ M'$) can be written as 
\begin{equation}
    \rho \ddot{\vect{\xi}} = \vect{f}_1[\vect{\xi}] + \rho \vect{a}_{\rm tide}. 
    \label{eq:eom_general}
\end{equation}
Here $\vect{\xi}=\vect{\xi}(\vect{r}, t)$ is the Lagrangian displacement of a fluid element at location $\vect{r}$ inside the primary and at time $t$, $\vect{f}_1$ characterizes the linear internal restoring force, and $\vect{a}_{\rm tide} = -\nabla U$ is the tidal acceleration. The tidal potential can be expanded as
\begin{equation}
U(\vect{r}, t) = - \! \sum_{l\geq 2, m} \! \! W_{lm}\frac{G  M'}{D(t)} \left[\frac{r}{D(t)}\right]^l \!\! Y_{lm}(\theta, \phi) {\rm e}^{-\imag \sigma t},
\label{eq:potential}
\end{equation}
where $Y_{lm}$ is the spherical harmonic function, $D$ is the orbital separation, and $\sigma = m(\Omega_{\rm orb}-\Omega_{\rm s})$ is the forcing frequency with $\Omega_{\rm orb}$ and $\Omega_{\rm s}$ being respectively the orbital frequency and spin frequency of the primary. The non-vanishing terms in the coefficient $W_{lm}$ are $W_{2\pm2}=\sqrt{3\pi/10}$ and $W_{20}=-\sqrt{\pi/5}$.

To solve for the internal response $\vect{f}_1$, we preform a six-dimension phase-space expansion (\citealt{Schenk:02, Weinberg:12}, hereafter, \citetalias{Weinberg:12})
\begin{equation}
   \begin{bmatrix} 
      \vect{\xi} (\vect{r}, t) \\
      \dot{\vect{\xi}}(\vect{r}, t)  \\
   \end{bmatrix}
=\sum q_a(t) 
  \begin{bmatrix} 
      \vect{\xi}_a (\vect{r}) \\
      -\imag \omega_a \vect{\xi}_a(\vect{r})  \\
   \end{bmatrix},
   \label{eq:expansion}
\end{equation}
where $q_a(t)$ is the amplitude and $\vect{\xi}_a(\vect{r})$ is the eigenvector of an eigenmode labeled by subscript $a$. The summation runs over all mode quantum numbers and both signs of eigenfrequency. 

For simplicity, we use the \emph{adiabatic} solutions to calculate the mode structure\footnote{This assumption is justified because the most-resonant mode when the binary has an orbital period in the range $10\,{\rm min}<P_{\rm orb} < 20\,{\rm min}$ will be weakly damped. The non-adiabatic equations will be solved only to related the temperature perturbation induced by each mode $\Delta T_a/T$ as a function of $\xi_a^r/R$.} and ignore all the rotational corrections except for the Doppler shift in the driving frequency.  Under these simplifications, the eigenvectors (solutions of $\vect{f}_1[\vect{\xi}_a] + \rho \omega_a^2\vect{\xi}_a = 0 $) can be further separated into radial and horizontal parts as 
\begin{equation}
\vect{\xi}_a (\vect{r}) = \left[\xi_a^r(r) \vect{e}_{r} + \xi_a^h(r) r \nabla \right] Y_{l_a m_a }(\theta, \phi),
\label{eq:vect_xi_decomp}
\end{equation}
with $\xi_a^r$ and $\xi_a^h$ both real. Furthermore, the set of eigenmodes form a complete, orthonormal basis. We choose the normalization of each mode such that 
\begin{equation}
2\omega_a^2 \int \diff^3 r \rho \vect{\xi}_{a}^\ast \cdot\vect{\xi}_{b}  = \frac{GM^2}{R} \delta_{ab} \equiv E_0 \delta_{ab}, 
\label{eq:xi_norm}
\end{equation}
and a mode with unity amplitude and its complex conjugate together will thus have an energy $E_0$.

Equation~(\ref{eq:eom_general}) can now be cast into a set of ordinary differential equations describing each mode's amplitude
\begin{gather}
    \dot{q}_a + (\imag\omega_a + \gamma_a)q_a = \imag \omega_a U_a,
    \label{eq:amp_a}
\end{gather}
where we have introduced a damping term $\gamma_a$ to account for the non-adiabatic dissipation (including both thermal diffusion and turbulent damping), and 
\begin{equation}
    U_a =W_{lm} Q_{a} \left( \frac{ M'}{ M} \right)  \left(\frac{R}{D}\right)^{l+1} {\rm e}^{-\imag \sigma t}.
    \label{eq:U_a}
\end{equation}
Here we have defined the tidal overlap of mode $a$ as
\begin{equation}
    Q_{a} = \frac{1}{MR^l} \int \diff^3 r \rho \vect{\xi}^\ast \cdot \nabla \left(r^l Y_{lm}\right).\label{eq:Qa}
\end{equation}
Note that to conserve the angular momentum it requires $l=l_a$ and $m=m_a$. 

The steady-state solution of the amplitude equation (\ref{eq:amp_a}) can be obtained by setting $\dot{q}_a = -\imag \sigma q_a$, which leads to 
\begin{equation}
    q_a = \frac{-\omega_a}{ \Delta_a + \imag \gamma_a} U_a,
\end{equation}
where $\Delta_a \equiv \sigma - \omega_a$ is frequency detuning. Such a mode oscillates at an angular frequency $\sigma$ in the \emph{corotating} frame and at $m\Omega_{\rm orb}$ in the \emph{inertial} frame. 
For future convenience, we will further define $\phi_a$ as the excess phase relative to the driving, 
\begin{equation}
    \phi_a \equiv \angle\left[\frac{-\omega_a}{ \Delta_a + \imag \gamma_a}\right].
    \label{eq:phi_a}
\end{equation}
The mode energy is given by (including also the contribution from the complex conjugate)
\begin{equation}
    E_a= q_a^\ast q_a E_0 = \frac{\omega_a^2}{\Delta_a^2+\gamma_a^2} |U_a|^2 E_0. 
    \label{eq:E_a}
\end{equation}

We can see that the dynamical tide is dominated by the most resonant mode with smallest $|\Delta_a|$. As shown by various previous studies (see, e.g., \citealt{Fuller:12a}; \citealt{Burkart:13}, hereafter, \citetalias{Burkart:13}; \citetalias{Yu:20}), this component dominates the tidal dissipation and the spin evolution of the WD. Specifically, the energy dissipation rate can be written as\footnote{Note that in this study, $E_a$ contains already the contributions from a mode and its complex conjugate. This is different from the convention used in \citetalias{Yu:20} where $E_a$ stands for the energy of a mode alone.}
\begin{equation}
    \dot{E}_{\rm diss} \simeq 2 \gamma_a E_a.
    \label{eq:dE_diss}
\end{equation}
In the above equation, mode $a$ corresponds to the most resonant mode with respect to the tidal forcing as it dominates the total energy dissipation. The tidal torque and the total tidal energy transfer rate (both from the orbit to the WD) are further related to the dissipation rate by (see, e.g., \citetalias{Yu:20})
\begin{align}
    &\tau_{\rm tide} = \frac{m}{\sigma} \dot{E}_{\rm diss}, \label{eq:tau_tide}\\
    &\dot{E}_{\rm tide} = \frac{m\Omega_{\rm orb}}{\sigma} \dot{E}_{\rm diss}. 
\end{align}
If we further assume that the WD maintains solid-body rotation, then its spin evolves as 
\begin{equation}
    \dot{\Omega}_{\rm s} = \frac{\tau_{\rm tide}}{I_{\rm WD}}.
    \label{eq:dOmega_s}
\end{equation}
The tidal evolution is closed once we further include the orbital evolution due to gravitational-wave radiation,
\begin{equation}
\dot{\Omega}_{\rm orb} \simeq \frac{96}{5} \left(\frac{G\mathcal{M}_c}{c^3}\right)^{5/3} \Omega_{\rm orb}^{11/3},
\label{eq:dOmega_gw}
\end{equation}
where $\mathcal{M}_c = (M M')^{3/5}/(M+M')^{1/5}$ is the chirp mass of the binary, and we have dropped the tidal back-reaction on the orbit since it is only a small correction to the orbital evolution in the range of interest. By simultaneously evolving Equations (\ref{eq:dOmega_s}) and (\ref{eq:dOmega_gw}), one can then obtain the evolution of the forcing frequency $\sigma$ and thus the tidal amplitude $q_a$. We will examine this in more details in Sec.~\ref{sec:tidal_evolution}. 

So far we have been focused on the dynamical tides that affect the spin evolution. Meanwhile, there is also an equilibrium component that accounts for the ellipsoidal variability. The equilibrium tide corresponds to setting the $\ddot{\vect{\xi}}$ term in Equation~(\ref{eq:eom_general}) to zero. If one further ignores the variation in the gravitational potential induced by the perturbed fluid (i.e., adopting the Cowling approximation), then the total equilibrium tide displacement, after doing the same decomposition as in Equation~(\ref{eq:vect_xi_decomp}),  can be written as 
\begin{equation}
    \frac{\xi^r_{\rm eq}}{R}\Big{|}_{lm} \simeq  W_{lm} \frac{M'}{M} \left(\frac{r}{D}\right)^{l+1}.
    \label{eq:xi_r_eq}
\end{equation} 
The temporal variation of this component is $\exp\left(-\imag\sigma t\right)$ in the \emph{corotating} frame and as $\exp\left(-\imag m\Omega_{\rm orb} t\right)$ in the \emph{inertial} frame.

\begin{figure}
   \centering
   \includegraphics[width=0.45\textwidth]{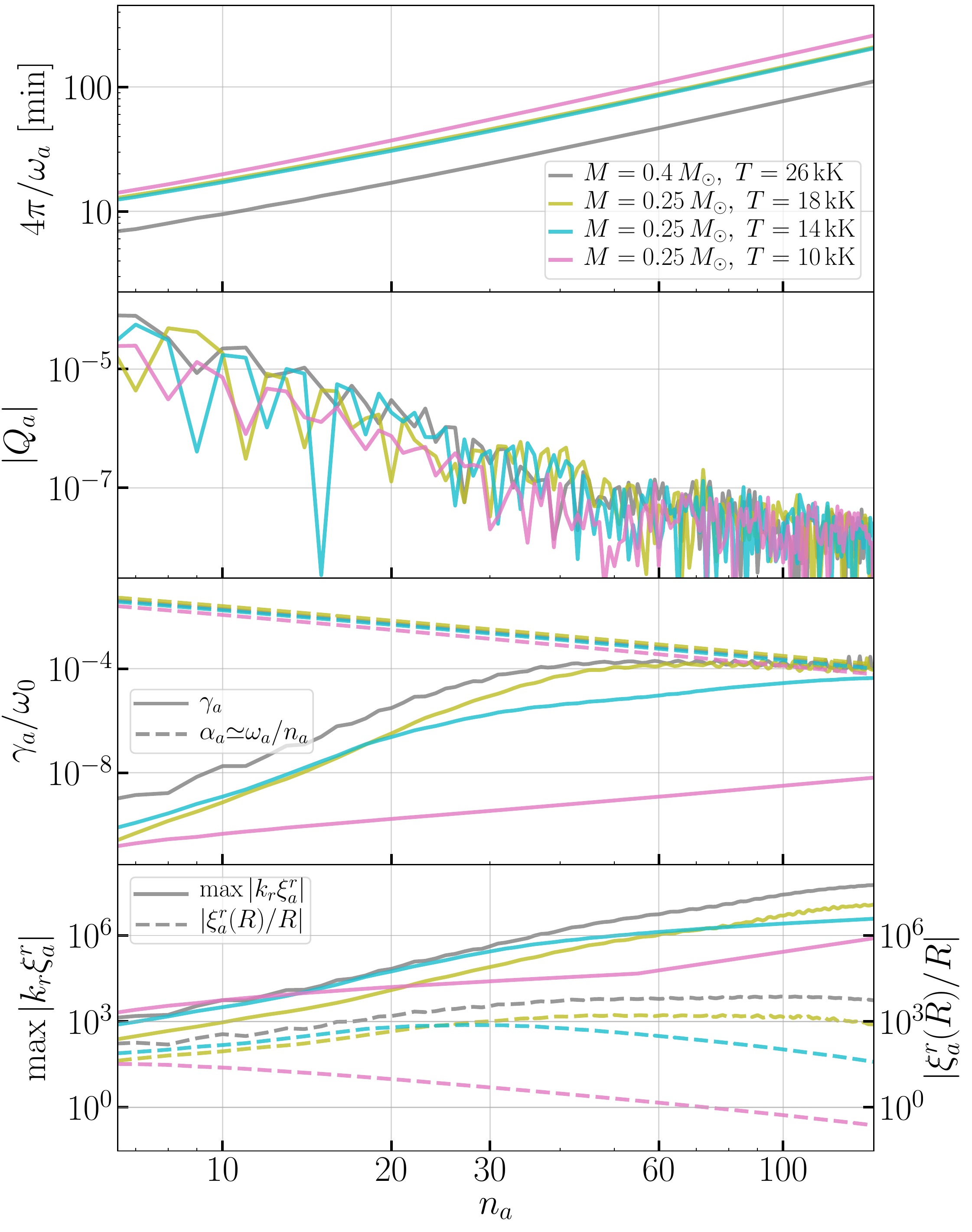} 
   \caption{From to top to bottom, we show respectively (twice) the mode's eigen-period $2P_a=4\pi/\omega_a$, the tidal overlap $|Q_a|$, the dissipation rate $\gamma_a$, and the maximum shear throughout the WD and the radial Lagrangian displacement at the surface at unity mode amplitude ($q_a=1$). All quantities are presented as a function of the radial order $n_a$ of a mode $a$. We use the color grey, olive, cyan, and pink to indicate quantities associated with the T26, T18, T14, and T10 Models, respectively. 
   }
   \label{fig:combo_lin_par}
\end{figure}

In Figure~\ref{fig:combo_lin_par} we show various mode parameters for the four WD models. We use the color (grey, olive, cyan, pink) to represent Model (T26, T18, T14, T10), respectively, and we will adopt the same coloring convention when comparing different models throughout the rest of the paper. The mode structures are obtained numerically from the stellar oscillation code  \texttt{GYRE}\citep{Townsend:13, Townsend:18}, and the detailed method of evaluating each quantity can be found in, e.g., \citetalias{Burkart:13, Yu:20}. 

For future convenience, we approximate the tidal overlap $Q_a$  (second panel; see also Equation~\ref{eq:Qa}) as 
\begin{equation}
    \ln Q_a(n_a) = x_1 \ln n_a + x_0,
    \label{eq:Qa_vs_na}
\end{equation}
for the rest of the calculation.
The values of fitting constants $x_1$ and $x_0$ are summarized in Table~\ref{tab:Qa}. All other parameters are directly taken from Figure~\ref{fig:combo_lin_par}.

\begin{table}
\begin{center}
\caption{\label{tab:Qa}Constants $x_1$ and $x_0$ for approximating the tidal overlap $Q_a$ according to Equation~(\ref{eq:Qa_vs_na})}
\begin{tabular}{ccccc}
    Model   & T26 & T18 & T14 & T10 \\
 \hline 
 $x_1$ & -3.1 & -2.5 & -3.0 & -2.7 \\
 $x_0$ & -4.2 & -6.6 & -5.1 & -6.6
\end{tabular}
\end{center}
\end{table}

A panel of particular importance is the third one where we show the dissipation rates due to radiative diffusion (including also electron conduction; solid traces).\footnote{For the hot WD models of interest, we find damping due to turbulent convection (see, e.g., \citetalias{Burkart:13}) is subdominant and can thus be ignored.} At large radial orders, it becomes comparable to the inverse group traveling time of a mode $\alpha_a$ (which is also the mode spacing between adjacent g-modes under the WKB approximation), defined as
\begin{equation}
    \alpha_a = \frac{2\pi}{t_{\rm grp}} = \frac{\pi}{\int (\diff k_r/\diff \omega_a) \diff r}\simeq \frac{\omega_a}{n_a}.
\end{equation}
For the hot T26 and T18 Models, we have $\gamma_a \sim \alpha_a$ for modes with $n_a\gtrsim 60$ and for the T14 Model it holds for modes with $n_a\gtrsim 200$. This means those modes are in fact traveling waves effectively. While our approximation of using the adiabatic mode structures may be inaccurate for them, we will show in Section~\ref{sec:tidal_evolution} that they barely affect the tidal evolution when the orbital period $P_{\rm orb}$ is less than 25 min. When $P_{\rm orb}\lesssim 25\,{\rm min}$, the most resonant mode will be a low-order, weakly damped mode that is in the standing-wave regime (apart from mode breaking which we will discuss shortly).

Because the high-order modes are in the traveling-wave regime, we do not expect the global parametric instability~\citepalias{Yu:20} to be significant for the hot models T26, T18, and T14. Instead, we will use the local wave-breaking criterion to estimate the onset of nonlinearity~(see, e.g., \citealt{Fuller:12a}; \citetalias{Burkart:13})
\begin{equation}
    2\max_{r} |q_a k_r \xi_a^r|= 1,\quad\text{(wave-breaking condition)},
    \label{eq:wave_break_cond}
\end{equation}
where $k_r \xi_a^r$ corresponds to the shear of a mode and we find the maximum of it over its propagation cavity. Note that the factor of 2 on the left-hand side accounts for the contributions to the shear from both mode $a$ and its complex conjugate.\footnote{Specifically, the complex conjugate of mode $a$ has an eigenfrequency of $-\omega_a$ and azimuthal quantum number $-m_a$. } 
For g-modes, the condition above corresponds to the perturbed buoyancy overturning the background stratification such that the mode oscillation can no longer be supported. On the other hand, if the shear is everywhere below unity, then a linear treatment of the problem should be justified (but see Section~\ref{sec:nonlinearity}). 

In the bottom panel of Figure~\ref{fig:combo_lin_par} we show the shear at unity mode amplitude ($q_a=1$) in the solid traces. Whereas in a cold WD like T10 the shear scales as $n_a$ because the modes all have essentially the same outer boundary where the shear reaches the maximum\footnote{This is true for modes with $n_a<60$; modes with higher radial order in fact reaches the maximal shear at the inner boundary for the T10 model.} (see also \citetalias{Yu:20}), we see a much sharper power-law dependence with respect to $n_a$ in the hot models as g-modes have different outer boundaries in these models (see Figure~\ref{fig:combo_prop}). 

Another quantity of interest is the Lagrangian displacement at the surface which we show in the dashed traces in the bottom panel. At the same mode amplitude, $|\xi_a^r(R)/R|$ increases with increasing temperature. This again stems from the fact that g-mode in a hotter WD can propagate closer to the surface than in a colder one and hence experience smaller exponential decay in the convective zone.

\subsection{Flux variation}
\label{sec:flux_eqs}
Once we obtained the value of $\xi^r_a$ based on the formalism in Section~\ref{sec:tide_eqs}, we then use \texttt{GYRE}~\citep{Townsend:13, Townsend:18} to solve the non-adiabatic fluid equations to estimate the flux variations. The quantity we are particularly interested in is the perturbation on the effective temperature $\Delta T$ which, for each mode, can be related to $\xi^r_a$ as (see also \citealt{Dupret:03})
\begin{align}
    &\frac{\Delta T_a }{T}(t, \theta, \phi) = f_{T, a} q_a(t) \frac{\xi_a^r(R)}{R} Y_{lm}(\theta, \phi) {\rm e}^{\imag \psi_{T, a}},\label{eq:dT_from_xi}
\end{align}
where 
the coefficients $f_{T,a}$ and $\psi_{T,a}$ are real constants for each mode and can be obtained from \texttt{GYRE}. 

With $\Delta T$ and $\xi^r$, our goal is to obtain variation in the emergent flux whose spectral density is formally defined as
\begin{align}
    &H_{\lambda}(\tau{=}0, t; T, g) =\frac{1}{4\pi} \int\int I_\lambda (\tau{=}0, \mu, \phi, t; T, g) \mu \diff\mu \diff\phi,
    \label{eq:H_lambda}
\end{align}
with $I_\lambda$ the specific intensity and $\mu = \cos\theta$. We further define $H=\int H_\lambda d\lambda$ as the integrated flux. 

If we denote $H^{(0)}$ as the unperturbed flux and $H^{(1)}_a$ its linear perturbation due to mode $a$, we have (see, e.g., \citealt{Pfahl:08, Burkart:12, Fuller:17})
\begin{align}
    \frac{H^{(1)}_a}{H^{(0)}}(t) &=  \left[(2b_l -c_l) + 4\beta^{(1)}(T)b_lf_{T, a}\e^{\imag \psi_{T, a}}\right] \nonumber \\
    &\times  q_a(t)\frac{\xi_a^r(R)}{R} Y_{lm}(\Theta_0, \Phi_0) 
    \label{eq:lin_flux_var}
\end{align}
where the first term in the bracket corresponds to the flux variation induced by change in the surface area, and the second term corresponds to that due to temperature fluctuation. In the above equation, $(\Theta_0, \Phi_0)$ is the polar coordinates of the line of sight in the WD frame. The coefficients $b_l$ and $c_l$ are disc-integral factors accounting for the $\mu$ dependence in the specific intensity, 
\begin{align}
    &b_l = \int_0^l \mu P_l(\mu) h(\mu)\diff \mu, \label{eq:b_l}\\
    &c_l = \int_0^1 \left[2\mu^2\frac{\diff P_l(\mu)}{\diff \mu } - (\mu-\mu^3)\frac{\diff^2P_l(\mu)}{\diff \mu^2}\right]\diff \mu,
\end{align}
with $h(\mu)$ the limb darkening function normalized as $\int_0^1 \mu h(\mu) \diff \mu =1$. Here we assume Eddington limb darkening for simplicity with $h(\mu)=1+3\mu/2$, and numerical values of $b_l$ and $c_l$ are computed in \cite{Burkart:12} for $l=0-5$. Lastly, the coefficient $\beta^{(1)}(T)$ accounts for the sensitivity of a telescope, 
\begin{align}
    \beta^{(1)}(T) = \frac{\int W(\lambda)\left[\partial H_\lambda^{(0)}/\partial \ln T\right] \diff \lambda}{4\int W(\lambda) H_\lambda^{(0)} \diff \lambda},
    \label{eq:beta_1}
\end{align}
where $W(\lambda)$ is the filter transmission. Note that when $W=1$ and the integration is preformed over all the wavelengths, we have $\beta^{(1)}=1$ for blackbody radiation. For ZTF, it is sensitive to around $400-900\,{\rm nm}$. However, at shorter wavelengths the sensitivity is slightly worse than at longer wavelengths~\citep{Dekany:20}. We thus approximate 
\begin{equation}
    W(\lambda) = 
    \begin{cases}
    1\text{, if $450\,{\rm nm}<\lambda < 900\,{\rm nm}$,}\\
    0\text{, elsewhere}.
    \end{cases}
\end{equation}
We further assume $H_\lambda^{(0)}\propto B_\lambda$ the Planck spectrum. The resultant $\beta^{(1)}(T)$ are summarized in Table~\ref{tab:beta}. Note that $\beta^{(1)}(T)$ decreases as the temperature increases, because for the range of temperature of interest, $\partial B_\lambda /\partial \ln T$ peaks at a lower wavelength than $B_\lambda$, and they both peak at wavelengths shorter than $450\,{\rm nm}$. 

\begin{table}
\begin{center}
\caption{\label{tab:beta}Values of $\beta^{(1)}(T)$ and $\beta^{(2)}(T)$ as a function of $T$ (see Equations~\ref{eq:beta_1} and \ref{eq:beta_2}). Here we assume the unperturbed flux $H_\lambda^{(0)}\propto B_\lambda$ the Planck spectrum, and a top-hat filter $W(\lambda)$ transmitting all the photons in the $450\,{\rm nm}<\lambda < 900\,{\rm nm}$ range.}
\begin{tabular}{ccccc}
$T$ [kK]   & $26$ & $18$ & $14$ & $10$ \\
\hline
$\beta^{(1)}(T)$ & 0.39 & 0.46 & 0.54 & 0.67 \\
$\beta^{(2)}(T)$ & 0.32 & 0.77 & 1.41 & 3.15 \\
\end{tabular}
\end{center}
\end{table}

\begin{figure}
   \centering
   \includegraphics[width=0.45\textwidth]{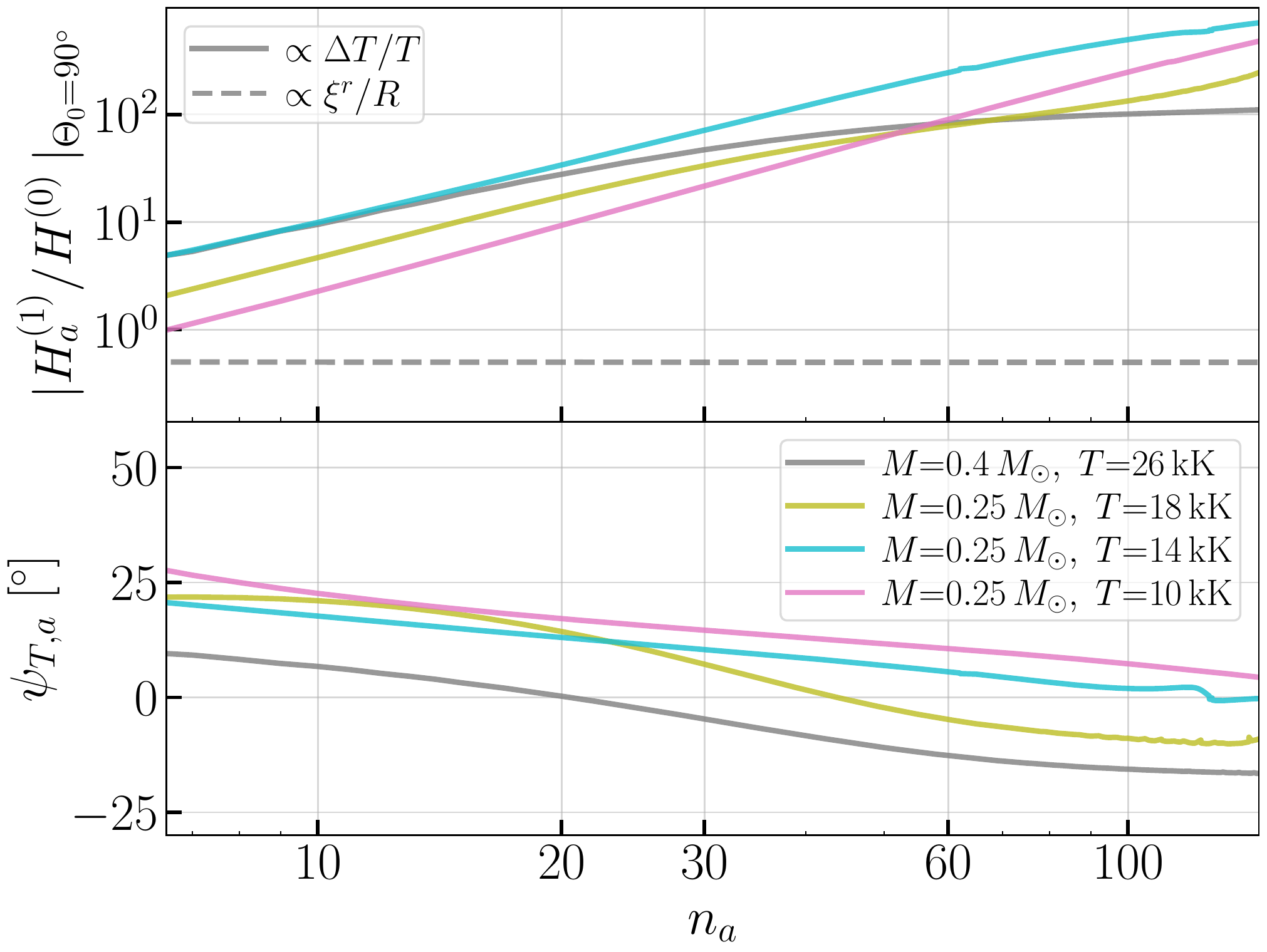} 
   \caption{Top panel: fractional flux variations as a function of modes' radial orders. When generating the plot, we have set each mode's amplitude such that the radial Lagrangian displacement at the surface is $q_a \xi_a^r(R)/R=1$ for the mode. We have also assumed an optimal viewing angle with inclination $\Theta_0=90^{\circ}$. For high-order g-modes, the total flux variation is dominated by the temperature fluctuation (solid trace) instead of the change in the surface area (dashed trace).
   Bottom panel: the excess phase $\psi_{T, a}$ of the temperature fluctuation relative to the surface displacement (Eqaution~\ref{eq:dT_from_xi}).
   For modes with $20<n_a<60$, we find $|\psi_T|\lesssim 10^\circ$. As a result, the observed phase of flux variation is dominated by $\phi_a$ (Equation~\ref{eq:phi_a}). }
   \label{fig:combo_flux_fix_xir}
\end{figure}

We are now ready to relate the surface Lagrangian displacement of a mode at the surface $\xi_a^r(R)/R$ to the fractional flux variation by utilizing Equation~(\ref{eq:lin_flux_var}). One example is shown in Figure~\ref{fig:combo_flux_fix_xir} where we show in the top panel the linear flux due to each mode if we set its amplitude such that $q_a\xi_a^r(R)/R=1$. We use solid traces to represent the flux particularly due to the temperature fluctuations (the second term in the bracket in Equation~\ref{eq:lin_flux_var}) and dashed traces that due to the surface area variations (the first term in Equation~\ref{eq:lin_flux_var}; note it is the same for different models because we fixed the surface displacement in this plot). We see that the temperature effect increases with increasing radial order and it typically dominates over the area effect by  orders of amplitude for high order g-modes with $n_a>20$ that are responsible to the dynamical tide. As we will see shortly, this is in contrast with the equilibrium tide (largely due to the f-mode and low-order g- and p-modes) where the two effects have comparable contributions to the flux. 


In the bottom panel of Figure~\ref{fig:combo_flux_fix_xir} we also show the excess phase of the temperature oscillation relative to the displacement, $\psi_{T, a}$. For g-modes with $20<n_a<60$, we find typically $\psi_{T, a}\in (-10^\circ, 10^\circ)$. Therefore, the dynamical-tide flux variation is out of phase with the orbital drive by $\phi_a + \psi_{T, a}\neq0 \,\text{(or $180^\circ$)}$ in general (note $\phi_a$ is defined in Equation~\ref{eq:phi_a}). As we will see shortly, this is in contrast to the ellipsoidal variability induced by equilibrium tide, which has a phase of exactly $180^\circ$ relative to the drive (see Equation~\ref{eq:ellip_var}). Therefore, the excess pulsation phase is one of the cleanest signatures of the dynamical tide. 

\begin{figure}
  \centering
  \includegraphics[width=0.45\textwidth]{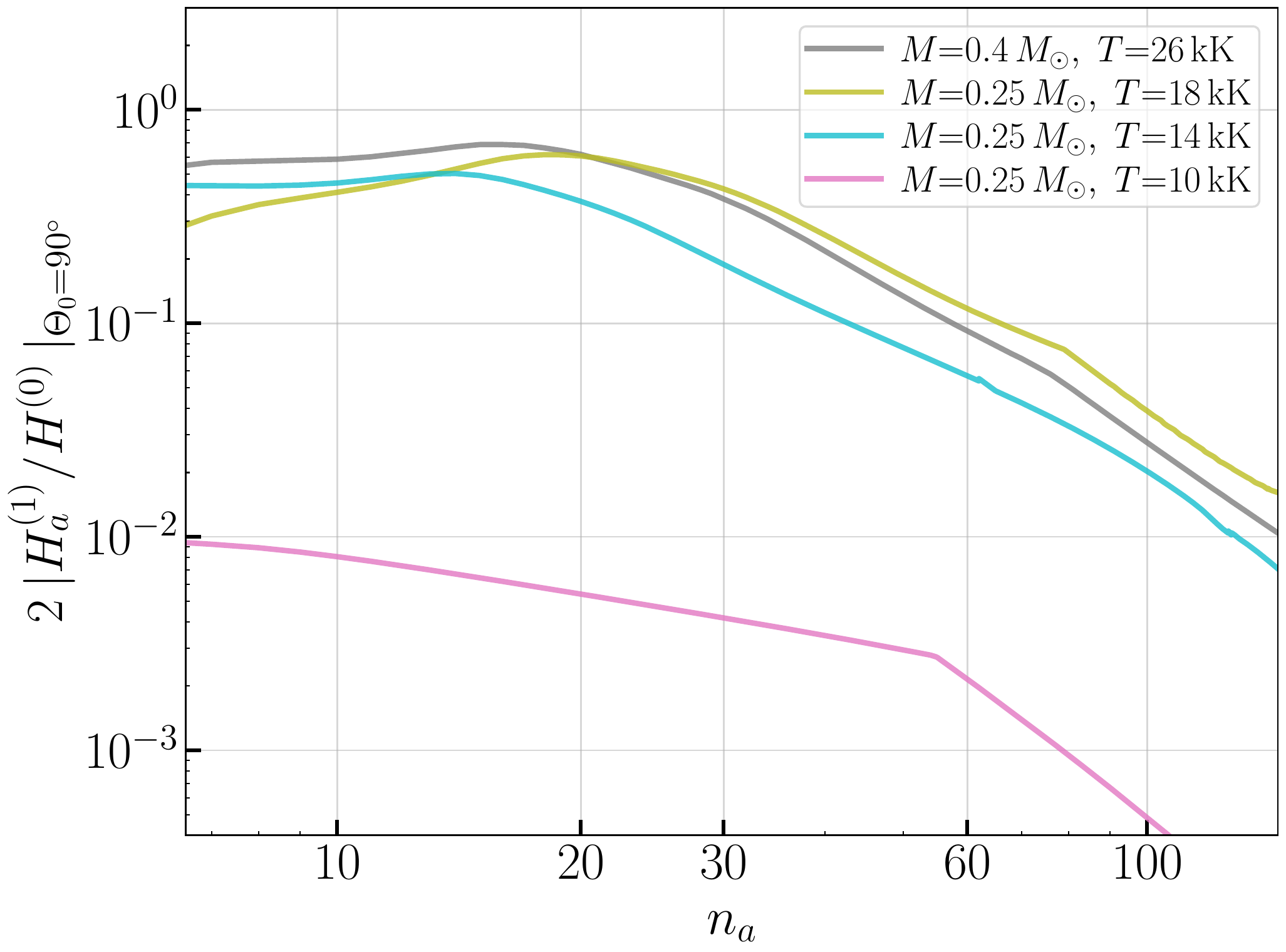} 
  \caption{The physical flux of a mode and its complex conjugate when the the mode amplitude reaches the wave-breaking condition (Equation~\ref{eq:wave_break_cond}). A mode with $n_a\lesssim 100$ in a hot WD with $T\gtrsim 14\,{\rm kK}$ can produce a significant flux variation ($>1\%$) before it breaks into a traveling wave. In contrast, no mode in the cold WD model T10 could perturb the flux by more than $1\%$ even at the wave-breaking amplitude. }
  \label{fig:combo_flux_brk}
\end{figure}

Furthermore, we may ask what is the maximum flux modulation can be produced by each mode. The result is shown in Figure~\ref{fig:combo_flux_brk}. Here we consider the flux variation at the wave-breaking amplitude (Equation~\ref{eq:wave_break_cond}; note the contributions from a mode's complex conjugate are included in both the shear and flux). For the three hot WD models T26, T18, and T14 (corresponding respectively to the grey, olive, and cyan traces in the plot), we note that modes with $n_a\lesssim 100$ can all produce more than $1\%$ perturbation to the flux before the mode breaks into a traveling wave. The coolest T10 mode, however, does not have any $n_a\gtrsim10$ mode that can modulate the flux by $1\%$ even at the wave-breaking amplitude. Indeed, because T10 has a propagation cavity whose outer boundary is the furthest away from the surface (see Figure~\ref{fig:combo_prop}), its g-modes are severely attenuated before reaching the surface. As a result, the dynamical tide has a smaller effect on the flux emerging from the surface in the T10 model. We will thus exclude T10 in our subsequent sections studying the tidal evolution and focus solely on the T26, T18, and T14 models. 


To conclude our discussions on the relation between tidal flux and mechanical displacement, we note that Equation~(\ref{eq:lin_flux_var}) applies to not only the dynamical tides but also equilibrium tide (changing the subscript ``$a$'' for a specific mode to ``eq''). For the equilibrium tide, the temperature fluctuation is further related to the Lagrangian displacement at the surface by 
\begin{equation}
    f_{T, {\rm eq}, lm} = -\frac{(l+2)}{4} \frac{\xi_{\rm eq}^r(R)}{R}\Big{|}_{lm}\simeq-\frac{(l+2)W_{lm}}{4}\frac{M'}{M}\left(\frac{R}{D}\right)^{l+1}.
\end{equation}
Therefore, the flux variation due to the equilibrium tide (i.e., the ellipsoidal variability) can be written as (in the inertial frame)
\begin{align}
    \frac{H^{(1)}_{{\rm eq}, lm}}{H^{(0)}}(t) &=  \left\{[2-\beta^{(1)}(T)(l+2)]b_l -c_l \right\}\nonumber \\
    &\times W_{lm} \frac{M'}{M} \left(\frac{R}{D}\right)^{l+1}Y_{lm}(\Theta_0, \Phi_0) \e^{-\imag m \Omega_{\rm orb}t}.
    \label{eq:ellip_var}
\end{align}
It is interesting to note that whereas the flux modulation due to dynamical tides is strongly dominated by the temperature fluctuation, that due to the equilibrium tide has similar contributions from both the temperature and the surface area variations. Specifically, for the dominant $l = m =2$ component with $\beta^{(1)}(T)\simeq 0.5$, we have $[2-\beta^{(1)}(T)(l+2)]\simeq 0$ and the ellipsoidal variability is mostly due to the term $\propto (-c_l)$. Furthermore, since $c_2= 39/20>0$, we see that the ellipsoidal variability has an exactly opposite sign (i.e., $180^\circ$ out of phase) relative to the tidal forcing $\sim \exp\left[-\imag 2 \Omega_{\rm orb}\right]$ (up to the phase due to the viewing angle, $\Phi_0$).

\section{TIDAL EVOLUTION}
\label{sec:tidal_evolution}

\subsection{Resonant lock}
\label{sec:res_lock}
The qualitative behavior of tidal evolution can be understood from a timescale argument. We can define
\begin{align}
    &T_{\rm s}(\Omega_{\rm orb}, \Omega_{\rm s}) = \frac{\Omega_{\rm orb}}{\dot{\Omega}_{\rm s}(\Omega_{\rm orb}, \Omega_{\rm s})},\\
    &T_{\rm gw}(\Omega_{\rm orb}, \Omega_{\rm s}) = \frac{\Omega_{\rm orb}}{\dot{\Omega}_{\rm orb}(\Omega_{\rm orb}, \Omega_{\rm s})},
\end{align}
as the spin-up and the orbital decay timescales, respectively. If $T_{\rm s} \gg T_{\rm gw}$, the tidal torque is too weak and the spin of the WD changes little as the orbit shrinks due to GW radiation. In the other limit with $T_{\rm s} \ll T_{\rm gw}$, the WD can be quickly spun-up by the tide until the synchronization condition\footnote{In principle, the right-hand-side of Equation~(\ref{eq:sync_cond}) should contain also terms like $(\partial \omega_a/\partial T) \dot{T}$, $(\partial \omega_a/\partial \Omega_{\rm s}) \dot{\Omega}_s$, and $(\partial \omega_a/\partial E_a) \dot{E}_a$, etc, corresponding respectively to shifts in a mode's eigenfrequency due to intrinsic cooling/tidal heating, rotation, and nonlinear effects, etc. For simplicity, we ignore these effects in our study here. }
\begin{equation}
    \dot{\Omega}_{\rm orb} - \dot{\Omega}_{\rm s} = 0,
    \label{eq:sync_cond}
\end{equation}
is met. We further define $P_{\rm c}$ as the critical orbital period at which Equation~(\ref{eq:sync_cond}) is first satisfied in the evolution.

\begin{figure}
   \centering
   \includegraphics[width=0.45\textwidth]{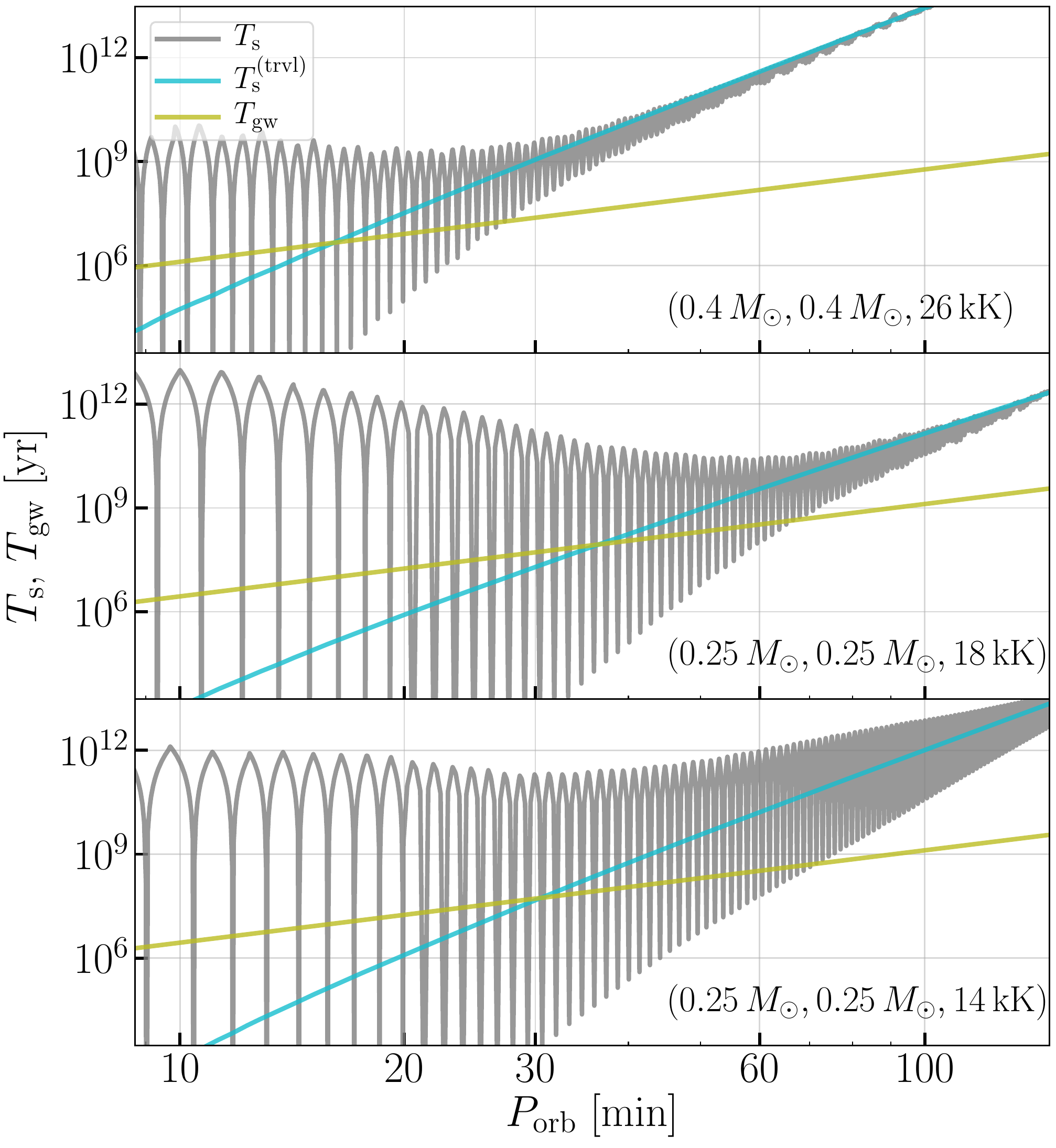} 
   \caption{Comparison of the spin-up timescale ($T_{\rm s}$; grey traces are for the standing-wave treatment and the cyan ones for the traveling-wave approximation) and the orbital decay timescale ($T_{\rm gw}$; olive trace) for different models. For top to bottom, the primary in each panel corresponds to respectively the T26, T18, and T14 models. The WD is assumed to be non-spinning. The companion's mass is the same mass as the primary, $M'=M$. }
   \label{fig:combo_timescales}
\end{figure}

In Figure~\ref{fig:combo_timescales} we compare the spin-up timescale with the orbital decay timescale for different WD models (from top to bottom, we have T26, T18, and T14, respectively; the companion is assumed to have a mass $M'=M$). Here we have fixed the WD to be non-rotating. At long orbital periods, the most-resonant mode corresponds to a high-order g-mode that is significantly damped by radiative diffusion and it is in fact a traveling wave (see Figure~\ref{fig:combo_lin_par}). In this regime, $T_{\rm s}$ is essentially a straight line with small spreading up and down.  We have $T_{\rm s}\gg T_{\rm gw}$ for all the models we consider and the WD cannot be synchronized in this traveling-wave regime. As the system evolves to shorter orbital periods, the most-resonant mode's radial order decreases (wavelength increases) and eventually its dissipation rate becomes smaller than the inverse group traveling time. Now the mode becomes a standing wave.\footnote{Since the tidal evolution is mostly due to a mode in the standing-wave regime with $\gamma_a \ll \alpha_a$, we use the adiabatic solution to estimate the mode's structure ($Q_a$, $\xi_a^r$, etc.) in this Section. } In this regime, $T_{\rm s}$ is characterized by a series of sharp dips, corresponding to the resonances of a series of modes. It is easy to see that Equation~(\ref{eq:sync_cond}) will be first satisfied at such a resonance. 

For the hottest model T26 with the most significant dissipation, the transition from the traveling-wave regime to the standing-wave regime does not happen until $P_{\rm orb}\lesssim 40\,{\rm min}$, and it thus has the smallest $P_{\rm c}\simeq 27\,{\rm min}$. As the WD's temperature decreases, the transition to the standing-wave regime happens at greater orbital period and the WD becomes synchronized with the orbit earlier on. For the T18 (T14) models, we have $P_{\rm c}\simeq 65\, (70)\,{\rm min}$. This trend of $P_{\rm c}$ increasing with decreasing $T$ can also be qualitatively understood by noticing that the tidal torque $\tau_{\rm tide}\propto 1/\gamma_a$ for a mode exactly on resonance. Since the dissipation rate is greater for a hotter WD model at a fixed radial order, the maximum (resonant) tidal torque is weaker. Thus a hotter model can achieve synchronization only with a lower-order mode (with shorter period) that both has smaller damping and overlaps more strongly with the tide.  

Unlike the temperature of the primary (i.e., $M$), the companion's mass plays a sub-dominant role in determining $P_{\rm c}$.  Changing $M'$ from $0.2\,M_\odot$ to $1.4\,M_\odot$, we find $P_{\rm c}$ changes only by $\mathcal{O}(1\%)$.  

Once synchronized, we would expect the synchronization to be maintained in the subsequent evolution $P_{\rm orb}<P_{\rm c}$. This corresponds to a resonance lock as discussed in \citetalias{Burkart:13} (see also \citealt{Burkart:14}). Indeed, the lock is possible because $T_{\rm s}\propto \Omega_{\rm orb}^{-3}$ whereas $T_{\rm gw}\propto\Omega_{\rm orb}^{-8/3}$ if Equation~(\ref{eq:sync_cond}) holds exactly, and therefore the tide can provide more than enough torque to keep the spin to co-evolving with the orbit. In reality, the lock happens at a slightly detuned location where $\Delta_a = m(\Omega_{\rm orb}-\Omega_{\rm s})-\omega_a < 0$. Consequently, the excess torque is compensated for by making the spin evolve slightly faster than the orbit, hence making the detuning $\Delta_a$ more negative with $|\Delta_a|$ being greater. 

In fact, if we combine the synchronization condition, Equation~(\ref{eq:sync_cond}), with Equations~(\ref{eq:dE_diss}), (\ref{eq:tau_tide}), and (\ref{eq:dOmega_s}), we can then solve for the post-synchronization mode amplitude as 
\begin{equation}
    |q_a|^2 \simeq \frac{\pi I_{\rm WD} \dot{\Omega}_{\rm orb}}{\gamma_a(P_{\rm c}) P_{\rm c} E_0},
\end{equation}
where $\gamma_a(P_{\rm c})$ is the dissipation rate of the most resonant mode at $P_{\rm c}$ with $4\pi/\omega_a\simeq P_{\rm c}$. Numerically, it can be expressed as 
\begin{align}
    &|q_a|^2\simeq 1.5\times10^{-13} \left(\frac{P_{\rm orb}}{20\,{\rm min}}\right)^{-11/3}\left(\frac{\mathcal{M}_{\rm c}}{0.22\,M_\odot}\right)^{5/3}\nonumber \\
    \times& \left(\frac{P_{\rm c}}{70\,{\rm min}}\right)^{-1} \left[\frac{\gamma_a(P_{\rm c})}{4\times 10^{-6}}\right]^{-1}\left(\frac{I_{\rm WD, 50}}{5.0}\right)\left(\frac{E_{0,48}}{6.7}\right)^{-1},
    \label{eq:mode_amp_post_sync}
\end{align}
where $I_{\rm WD} = I_{\rm WD, 50} \times 10^{50}\,{\rm g\,cm^{2}}$ and $E_{0}=E_{0, 48}\times 10^{48}\,{\rm erg}$. 

Once we know a mode's amplitude, we can then combine it with Figure~\ref{fig:combo_lin_par} to obtain, e.g.,  the Lagrangian displacement at the surface $\xi_a^r(R)$. For the hot WD models T26, T18, and T14, we find that the most resonant modes at $P_{\rm c}$ typically have $|\xi_a^r(R)/R|\simeq 10^3-10^4$ with unity amplitude $|q_a|=1$. Equation~(\ref{eq:mode_amp_post_sync}) thus suggests a typical $|\xi_a^r(R)/R|\sim 10^{-3}$ at $P_{\rm orb}=20\,{\rm min}$, and it increases as $P_{\rm orb}^{-11/6}$ as the orbit decays to shorter periods.

\subsection{Traveling-wave approximation}
\label{sec:tw}
The above description remains valid until nonlinear effects kick in. As we argued in Section~\ref{sec:tide_eqs}, the parametric instability \citepalias{Yu:20} may not be significant in the hot WD models we consider here thanks to the high dissipation rate of high-order modes. Therefore, we adopt the local wave-breaking criterion, Equation~(\ref{eq:wave_break_cond}), as the onset of nonlinear effects.

Once entering this strongly nonlinear regime, the tidal wave no more reflects at the outer boundary (where wave-breaking happens) but behaves like a one-way traveling wave. 
We thus follow \citetalias{Burkart:13} and approximate the traveling-wave effect by replacing the linear dissipation rate $\gamma_a$ with the wave's inverse group-traveling time $\alpha_a$. To determine the most resonant ``mode'' responsible for the dynamical tide, we first interpolate $\omega_a$ as a function of $n_a$ and then solve $\sigma = \omega_a(n_a)$ allowing $n_a$ to take on non-integer values. Indeed, in the traveling-wave limit there is no discrete set of standing waves formed and we thus approximate the continuous response to the tidal forcing frequency $\sigma$ by allowing $n_a$ to vary continuously. The other structural parameters are similarly approximated by evaluating at non-integer values of $n_a$ using interpolation functions obtained using the eigenmodes. This allows us to write the traveling-wave amplitude as
\begin{equation}
    q_a^{\rm (trvl)}(\sigma) = \imag \frac{\sigma}{\alpha_a(\sigma)} U_a(\sigma),
    \label{eq:amp_a_trvl}
\end{equation}
where a superscript ``trvl'' has been used to indicate a quantity is evaluated under the traveling-wave limit. Note that by construction $\Delta_a^{\rm (trvl)} \equiv 0$ and $q_a^{\rm (trvl)}$ is $90^\circ$ out of phase with respect to the orbital driving. 
The tidal torque effectively reduces to \citepalias{Burkart:13}
\begin{equation}
    \tau_{\rm tide}^{\rm (trvl)}(\sigma)=4\frac{\sigma}{\alpha_a(\sigma)}|U_a(\sigma)|^2E_0.
    \label{eq:tau_tide_trvl}
\end{equation}


We incorporate the traveling-wave limit in two ways. In the first one, we interpolate the torque as (\citetalias{Burkart:13})
\begin{equation}
    \tau_{\rm tide} = \frac{\tau_{\rm tide}^{\rm (stnd)} + \tau_{\rm tide}^{\rm (trvl)} \left(2\max |q_a k_r\xi_a^r |\right)^z}{1 + \left(2\max |q_a k_r\xi_a^r |\right)^z},
    \label{eq:torque_interpolation}
\end{equation}
where $\tau_{\rm tide}^{\rm (stnd)}$ is the standing wave tidal torque (Equation~\ref{eq:tau_tide}), and $z= 25$ an interpolation constant (whose value our results are insensitive to as along as $z\gg 1$). We remind the reader again that we use twice the maximum shear of a single mode to account for contributions to the physical shear from both the mode itself and its complex conjugate. The formula above allows us to continue the spin evolution even when the wave enters the wave-breaking regime, though the linear amplitude $q_a$ may not be valid anymore as the interpolation scheme is designed for the tidal torque only.\footnote{In Equation~(\ref{eq:torque_interpolation}), the amplitude $q_a$ is always calculated using the linear expression Equation~(\ref{eq:amp_a}) and the linear damping rate $\gamma_a$. If we also similarly interpolate the damping rate between $\gamma_a$ and $\alpha_a$, and use the relations given by Equations~(\ref{eq:dE_diss}) and (\ref{eq:tau_tide}), we then recover the traveling-wave energy $\left[\left(q_a^{\rm (trvl)}\right)^\ast q_a^{\rm (trvl)}\right]$ with $q_a^{\rm (trvl)}$ in Equation~(\ref{eq:amp_a_trvl}) in the highly nonlinear regime.} 

Alternatively, we will also consider a case where we use exclusively the traveling-wave limit to study the tidal evolution. This is conceptually similar to the situation studied by \citet{Fuller:12a} where the wave is strongly nonlinear and even an off-resonant mode can still break. This scenario applies if the wave in fact breaks at a much lower threshold than Equation~(\ref{eq:wave_break_cond}). It is also relevant for less compact WDs with greater radii than those we consider here, so that the synchronization may in fact happen even with a highly-damped parent mode (e.g., for the T18 model, this corresponds to the synchronization happening at $P_{\rm c}\gtrsim 100\,{\rm min}$). The traveling-wave spin-up timescale is shown in the cyan traces in Figure~\ref{fig:combo_timescales}.


\subsection{Numerical evolution trajectories}
\begin{figure}
   \centering
   \includegraphics[width=0.45\textwidth]{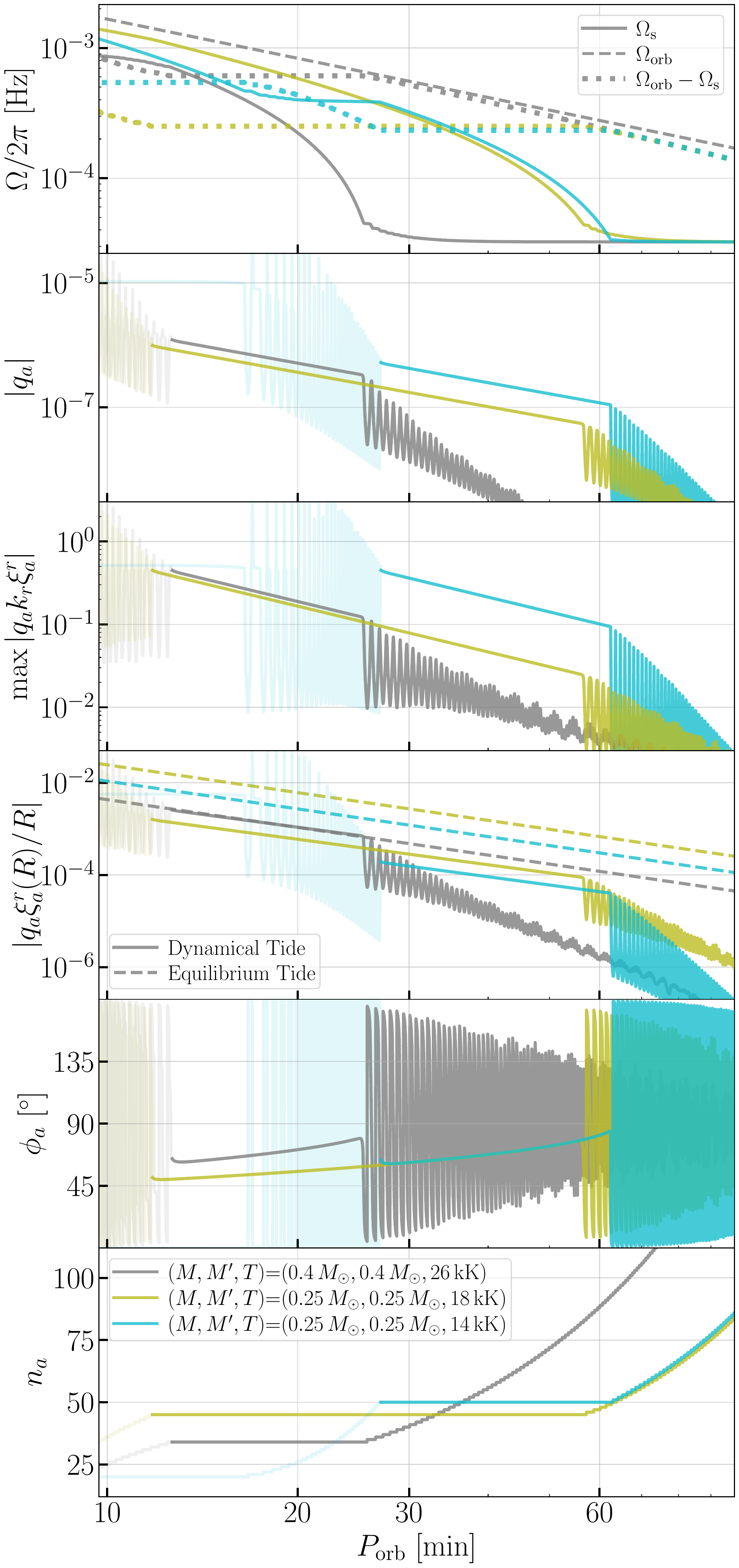} 
   \caption{Tidal evolution trajectories under the standing-wave assumption. From top to bottom, we show respectively the evolutionary trajectories of the orbital and spin frequencies, 
   the amplitude of the most resonant mode,
   the maximum shear throughout the WD, 
   the Lagrangian displacement at the stellar surface due to both dynamical  and equilibrium tide, 
   the phase of $q_a$ relative to the orbital forcing $\phi_a$,
   and the radial order of the most resonant mode. As a mode's amplitude increases with decreasing $P_{\rm orb}$, Equation~(\ref{eq:wave_break_cond}) is satisfied and the standing-wave picture is no more accurate. The trajectories in this regime are thus set to be transparent. 
   }
   \label{fig:combo_tidal_traj}
\end{figure}

We are now ready to examine the full numerical solutions to the tidal synchronization problem governed by Equations~(\ref{eq:dOmega_s}) and (\ref{eq:dOmega_gw}). We first assume the most resonant mode behaves as a standing wave until Eq.~\ref{eq:wave_break_cond} is satisfied. The result is shown in Figure~\ref{fig:combo_tidal_traj}. Here we use the color (grey, olive, cyan) to represent the (T26, T18, T14) models, respectively, and the companion mass is set to $M'=M$ in all the cases. The T10 model is excluded because it enters the wave-breaking regime at long orbital periods and even at the saturation amplitudes it cannot produce more than $1\%$ flux variation (Figure~\ref{fig:combo_flux_brk}). Our numerical integration starts at $P_{\rm orb} = 100\,{\rm min}$ and with an initial spin frequency $\Omega_{\rm s} = \Omega_{\rm orb}/5$. The final result is insensitive to the initial condition as long as the initial spin is small. 

In the top panel, we show the evolution of the spin frequency (solid traces), the orbital frequency (dashed traces), as well as their difference $\left(\Omega_{\rm orb} - \Omega_{\rm s}\right)$ (dotted traces). As we argued in Section~\ref{sec:res_lock}, at $P_{\rm orb}\gtrsim P_{\rm c}$ the spin frequency experience little evolution, whereas at $P_{\rm orb}\lesssim P_{\rm c}$ the asynchronicity $(\Omega_{\rm orb} - \Omega_{\rm s})$ becomes approximately a constant, indicating the WD is synchronized according to Equation~(\ref{eq:sync_cond}). Note the location of $P_{\rm c}$ is at a slightly smaller orbital period in the full numerical calculation than the estimation based on Figure~\ref{fig:combo_timescales}. This is because of the slight spin-up of the WD prior to synchronization. Nonetheless, the agreement is decent overall. 

We show the radial order of the most resonant mode in the bottom panel. Here we focus first on the dark traces assuming the tidal responses forms standing waves until Equation~(\ref{eq:wave_break_cond}) is satisfied. As a consequence of tidal synchronization, we see that $n_a$ becomes a constant when $P_{\rm orb}<P_{\rm c}$. For the (T26, T18, T14) models, we have respectively $n_a=(34, 45, 50)$. In other words, the system is resonantly locked to the mode with $n_a$. 

The post-synchronization amplitude of the most resonant mode is very well described by Equation~(\ref{eq:mode_amp_post_sync}) as shown in the second panel. Combining it with Figure~\ref{fig:combo_lin_par}, we can thus obtain the evolution of the maximum shear throughout the WD and the Lagrangian displacement at the surface, which are shown respectively in the third and forth panels of Figure~\ref{fig:combo_tidal_traj}.

From the third panel, the T14 model (cyan) shows a particularly strong shear compared to the other two models. This is because it has the smallest $\gamma_a(P_{\rm c})$ (see Equation~\ref{eq:mode_amp_post_sync} and Figure~\ref{fig:combo_lin_par}).  At $P_{\rm orb}\simeq 27\,{\rm min}$, we have $2\max|q_a k_r \xi_a^r|$ exceeding unity for the T14 model. The resonant lock is thus destroyed by wave breaking and the asynchronicity increases again due to the reduced tidal torque (Equation~\ref{eq:tau_tide_trvl}). The linear, standing-wave description is no more valid in this regime and we change the color to light-cyan in the plots. The the hotter T26 and T18 models, on the other hand, both have smaller shear and break when the binary becomes more compact. For the T26 model, this corresponds to $P_{\rm orb}\simeq 13\,{\rm min}$ and for T18, $P_{\rm orb}\simeq 12\,{\rm min}$ (see also Section~\ref{sec:discussion}). 

Whereas the T14 model has the greatest shear, its Lagrangian displacement at the surface is the smallest among all the models as shown in the forth panel. This can be understood by noticing the T14 model's propagation cavity does not extends to as close to the surface as the hotter models (Figure~\ref{fig:combo_prop}). Consistent with our discussions around Equation~(\ref{eq:mode_amp_post_sync}), we have $\xi_a^r(R)/R \sim 10^{-3}$ at $P_{\rm orb}\simeq 20\,{\rm min}$ for the T26 and T18 models. As comparisons, we also show the displacement induced by the $l=m=2$ equilibrium tide (Equation~\ref{eq:xi_r_eq}) in the dashed traces. Note that $\xi_{\rm eq}^r/R\propto \overline{\rho}^{-1}$ with $\overline{\rho}$ the mean density of the WD. For the densest T26 model, it has the smallest $\xi_{\rm eq}^r/R \simeq 10^{-3}$ at $P_{\rm orb}=20\,{\rm min}$ and is similar to that due to the dynamical tide. In contrast, the T18 model (with the smallest density) has a displacement due to the equilibrium tide that is about 10 times greater, $\xi_{\rm eq}^r/R\simeq 10^{-2}$. We further note that $\xi_{\rm eq}^r/R\propto P_{\rm orb}^{-2}$ whereas $\xi_{\rm a}^r/R\propto P_{\rm orb}^{-11/6}$ (see, Equations~\ref{eq:xi_r_eq} and \ref{eq:mode_amp_post_sync}). As a result, the ratio between the dynamical and equilibrium components stays nearly a constant for the post-synchronization part of the evolution.

Lastly, we show the phase $\phi_a$ (Equation~\ref{eq:phi_a}) in the fifth panel. At $P_{\rm c}$ we have $\phi_a\simeq 90^\circ$ as the synchronization is achieved at close to the resonance of a mode. In the subsequent evolution, $\phi_a$ decreases to smaller values whose reason is the following. The resonant lock only happens on the side with $\Delta_a < 0$. As we argued in Section~\ref{sec:res_lock}, the spin-up timescale decreases faster than the orbital decay timescale if a mode stays exactly on resonance, $T_{\rm s}\propto P_{\rm orb}^{3}$ whereas $T_{\rm orb} \propto P_{\rm orb}^{8/3}$, and thus the tidal torque is more than enough to maintain Equation~(\ref{eq:sync_cond}). This makes the asynchronicity decrease slightly and $\Delta_a$ become more negative. The excess tidal torque is therefore balanced by detuning the resonance. Consequently, $\phi_a$ decreases to smaller values.

\begin{figure}
   \centering
   \includegraphics[width=0.45\textwidth]{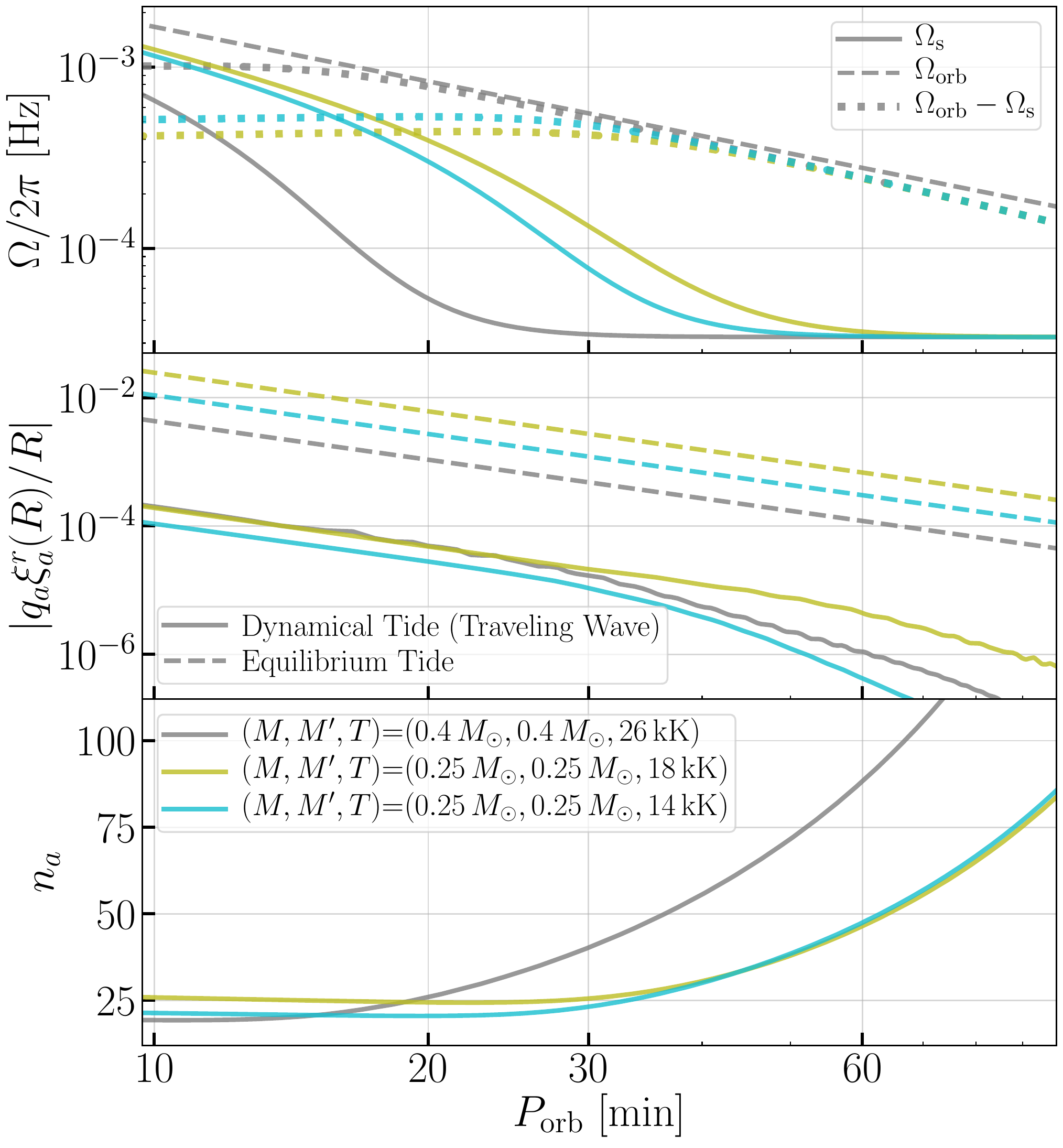} 
   \caption{Similar to Figure~\ref{fig:combo_tidal_traj} but now assumes the traveling-wave approximation (see Section~\ref{sec:tw}). 
   }
   \label{fig:combo_tidal_traj_trvl}
\end{figure}

The discussions above assumes the standing-wave picture applies until the physical shear reaching unity. For completeness, we also consider the tidal evolution trajectories based on the traveling-wave approximation in Figure~\ref{fig:combo_tidal_traj_trvl} in case the gravity waves break at a much  lower amplitude than our assumption in Equation~(\ref{eq:wave_break_cond}). It may also be appropriate if a shear layer forms and absorbs the gravity waves~\citep{Fuller:12b, Su:20}.  The spin evolution for this case is shown in the top panel of Figure~\ref{fig:combo_flux_traj_trvl} and in the bottom panel, we also show the value of $n_a$ obtained through $\sigma = 2(\Omega_{\rm orb}-\Omega_{\rm s})=\omega_a(n_a)$. Note $n_a$ now takes on non-integer values as discussed in Section~\ref{sec:tw}. As there are no more resonances, the amplitudes are greatly reduced in the traveling-wave limit compared to the standing wave case and so are the tidal torques. Consequently, the synchronization can happen only at shorter orbital periods. For T26, we have $P_{\rm c}\simeq 18\,{\rm min}$ and for T18 and T14, $P_{\rm c}\simeq 30\,{\rm min}$. Note that because of the orbital decay, an asynchronicity trajectory based on interpolating the standing-wave and traveling-wave torques (Equation~\ref{eq:torque_interpolation}; i.e., trajectories shown in Figure~\ref{fig:combo_tidal_traj}) quickly merges to that based on the traveling-wave approximation alone once the WD enters the strongly nonlinear regime. In other words, the result is insensitive to the spin history (i.e., the initial condition of the evolution). Although our calculation here overestimates the asynchronicity at long orbital periods, it should nonetheless give a decent approximation when the companion excites strongly nonlinear tidal waves. 

In the post-synchronous regime, the asynchronicity is not locked to the vicinity of a specific resonant mode as in the standing-wave case (\citetalias{Burkart:13}) but evolves slightly as we approximated the tidal torque as a smooth function with respect to the forcing frequency (see, e.g., \citealt{Fuller:12a}; \citetalias{Yu:20}). Nonetheless, Equation~(\ref{eq:mode_amp_post_sync}) still shows good agreement with the post-synchronous mode amplitude. Therefore, as shown in the middle panel, the ratio between the dynamical-tide displacement and the equilibrium-tide displacement stays approximately constant as the orbit decays even in the traveling-wave limit. 

The overall amplitude of the dynamical-tide displacement is significantly reduced relative to the standing-wave case due to both the reduced mode amplitude and the fact that the spin is more asynchronous with the orbit and a higher driving frequency excites tidal wave whose outer boundary is further away from the surface (Figure~\ref{fig:combo_prop}). The T26, T18, and T14 models all predict $|q_a\xi_a^r(R)/R|\simeq \text{a few}\times 10^{-5}$ at around $P_{\rm orb}\simeq 20\,{\rm min}$ in the traveling-wave limit. 

\section{Flux variation induced by dynamical tides}
\label{sec:flux_evolution}

Once we have determined the tidal evolution trajectories, we can then estimate the flux induced by the most resonant mode utilizing Equation~(\ref{eq:lin_flux_var}). 

\begin{figure}
   \centering
   \includegraphics[width=0.45\textwidth]{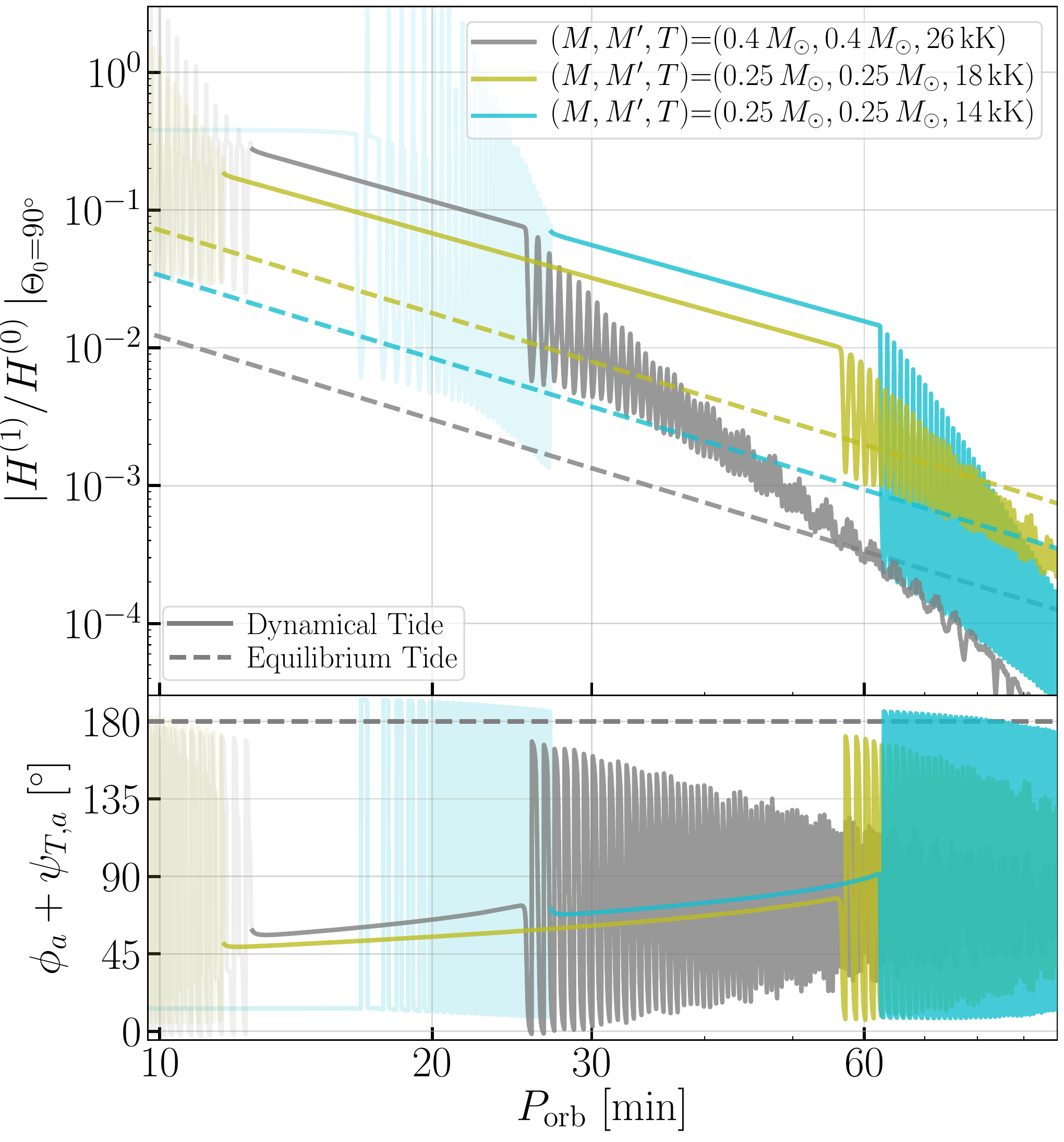} 
   \caption{Top panel: using the trajectory determined by Figure~\ref{fig:combo_flux_traj}, the evolution of the amplitude of the observed flux fluctuation due to dynamical tides (solid traces; including a mode and its complex conjugate). Also shown in the dashed  traces are amplitudes of flux variations due to equilibrium tides (i.e., the ellipsoidal variability).
   With an optimal viewing angle, the dynamical tide could induce a flux variation on the order of $10\%$ in the $10-25\,{\rm min}$ band and dominates over the ellipsoidal variation that varies at $2\Omega_{\rm orb}$. This should be a detectable signature in ground-based photometry. 
   Bottom panel: the excess phase of the flux variation relative to the orbital forcing. }
   \label{fig:combo_flux_traj}
\end{figure}

The result is shown in Figure~\ref{fig:combo_flux_traj} for the standing-wave calculation (Figure~\ref{fig:combo_tidal_traj}). In the top panel we show the expected flux variation amplitudes assuming an optimal viewing angle $\Theta_0=90^\circ$ (i.e., the binary is viewed edge-on). Here we have also included the contribution from a mode and its complex conjugate, $H^{(1)}=2{\rm Re}\left[H^{(1)}_a\right]$. The flux is exclusively due to the primary $M$ and we do not include the contributions from $M'$. The solid traces correspond to the flux due to dynamical tide (under the linear, standing-wave assumption), and as a comparison, we also show in the dashed traces the ellipsoidal variability due to the equilibrium tide.  
It is interesting to note that the dynamical tide can alter the observed flux by as much as $10\%$ at $P_{\rm orb}=20\,{\rm min}$ for T26 and T18. Even at relatively long orbital periods of $P_{\rm orb}\lesssim 60\,{\rm min}$, T18 and T14 could still produce $\mathcal{O}(1\%)$ photometric modulation due to dynamical tides. This could exceed the equilibrium tide's contribution by factor of about 3 (for T18) to about 30 (for T26 post synchronization). As we discussed in Section~\ref{sec:tidal_evolution}, this ratio should stay approximately a constant for the subsequent evolution (as long as the most-resonant g-mode does not break). The ratio also does not depend on the viewing angle because they both follow an angular dependence given by $Y_{22}(\Theta_0, \Phi_0)$. Therefore, our model predicts that the dynamical tide's pulsation should be directly observable in hot WDs with orbital periods in the $\lesssim 60$ minutes range. More importantly, since both the dynamical tide and the equilibrium tide varies at $2\Omega_{\rm orb}$ in the inertial frame, it would thus be critical to properly take into account the the dynamical tide's contribution if one wants to extract binary parameters based on the ellipsoidal variability. 

In the bottom panel we also show the phase of the oscillation (after factoring out the orbital phase and the viewing angle), which, for equilibrium tide, is $180^\circ$ as we discussed at the end of Section~\ref{sec:flux_eqs}, and for the dynamical tide, is essentially $\phi_a$ since $\psi_{T,a}$ is small for g-modes (around $\pm 10^\circ$; see Figures~\ref{fig:combo_tidal_traj} and \ref{fig:combo_flux_fix_xir}). Note that the predicted phase shift can be large ($45^\circ-90^\circ$) and easy to notice in the light curve. Consequently, the pulsation phase serves as a smoking gun signature of dynamical tides.

As for the ratio between the amplitudes of the dynamical tide's flux variation and the ellipsoidal variability, we find it depends most sensitively on the radius $R$. Whereas the equilibrium tide's flux perturbation simply increases as $R^3$, the dynamical tide typically has a more complicated dependence. Indeed, the radius affects both the WD's internal structure and the level of synchronization (consequently the most resonant mode). Various effects cancel each other to some extend and as a result, T26, T18 and T14 all produce similar levels of flux variation ($\mathcal{O}(10\%)$) due to dynamical tide while the equilibrium tide's flux various by more than an order of magnitude.

\begin{figure}
   \centering
   \includegraphics[width=0.45\textwidth]{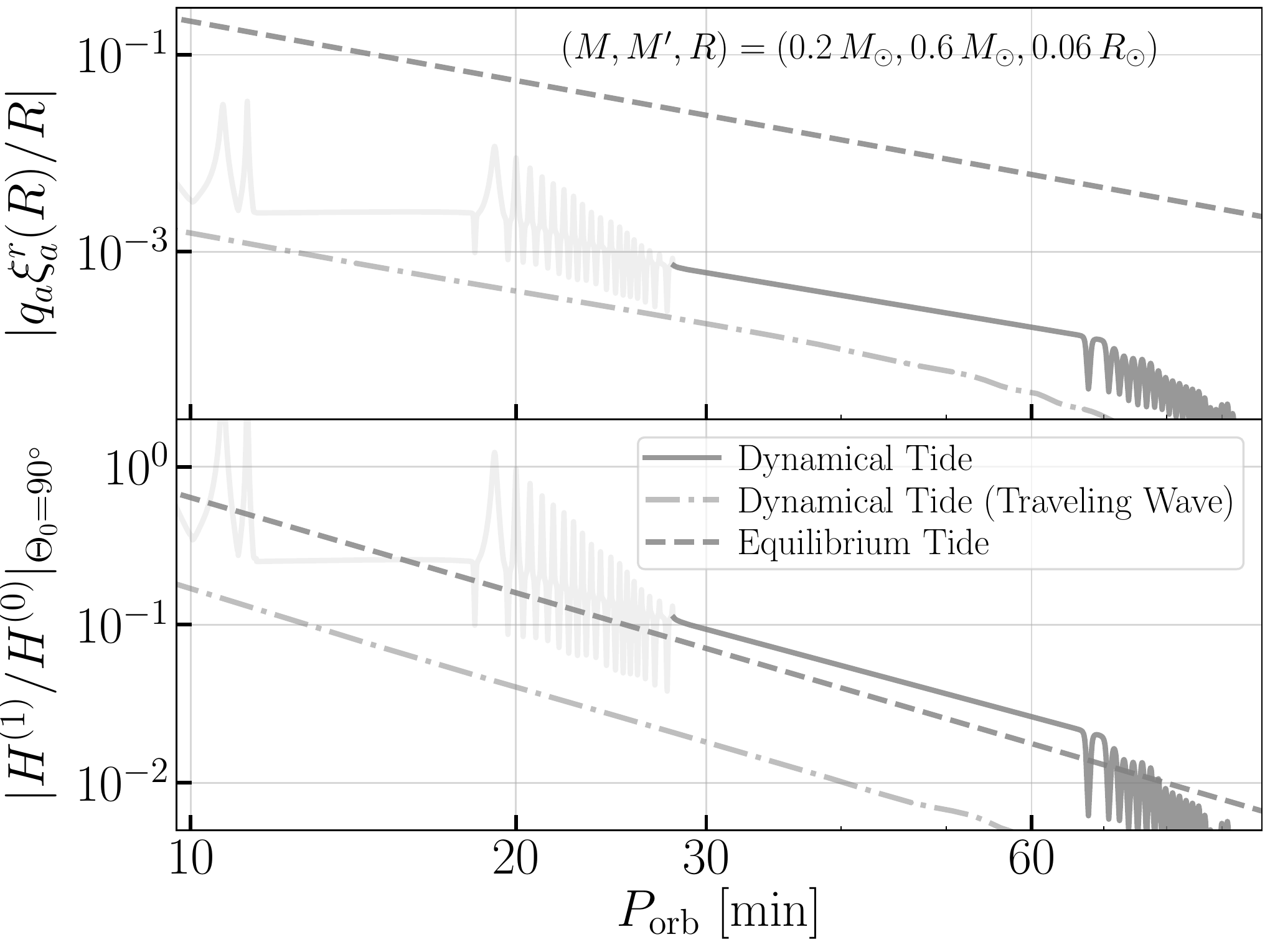} 
   \caption{Similar to Figure~\ref{fig:combo_flux_traj} but we rescaled the overall tidal amplitude according to $(M, M', R)=(0.2\,M_\odot, 0.6\,M_\odot, 0.06\,R_\odot)$. The other dimensionless internal parameters are the same as the T18 model and it should thus be representative for WDs having $T\simeq 18\,{\rm kK}$.
   The top panel shows the surface displacement of the most resonant mode and the bottom panel shows the flux variation (including both the mode shown in the top panel and its complex conjugate). The solid traces assume the tide excites a standing wave inside the WD, which breaks down when $P_{\rm orb}\lesssim 28\,{\rm min}$ and therefore we changed its color to light grey. As a comparison, we also show the displacement and the flux modulation under the traveling-wave limit in the dash-dotted lines. }
   \label{fig:M0p2_R0p06_flux_traj}
\end{figure}

We illustrate this point further in Figure~\ref{fig:M0p2_R0p06_flux_traj}. Here we consider a system with $(M, M', R)=(0.2\,M_\odot, 0.6\,M_\odot, 0.06\,R_\odot)$. We use these parameters to rescale the overall tidal strength while leaving the other dimensionless quantities the same as the T18 model. It should thus be representative for more extended WDs with temperature $T\simeq 18\,{\rm kK}$. As the radius increases, the equilibrium tide now perturbs the flux by about the same amount as the dynamical tide (in the standing-wave regime) whose flux increases only mildly. In fact, as the radius increases, $P_{\rm c}$ increases as well and the most resonant mode in this case has a higher radial order than the T18 model. This mode is more prone to wave-breaking, which happens at $P_{\rm orb}\simeq 28\,{\rm min}$ in this case.

\begin{figure}
   \centering
   \includegraphics[width=0.45\textwidth]{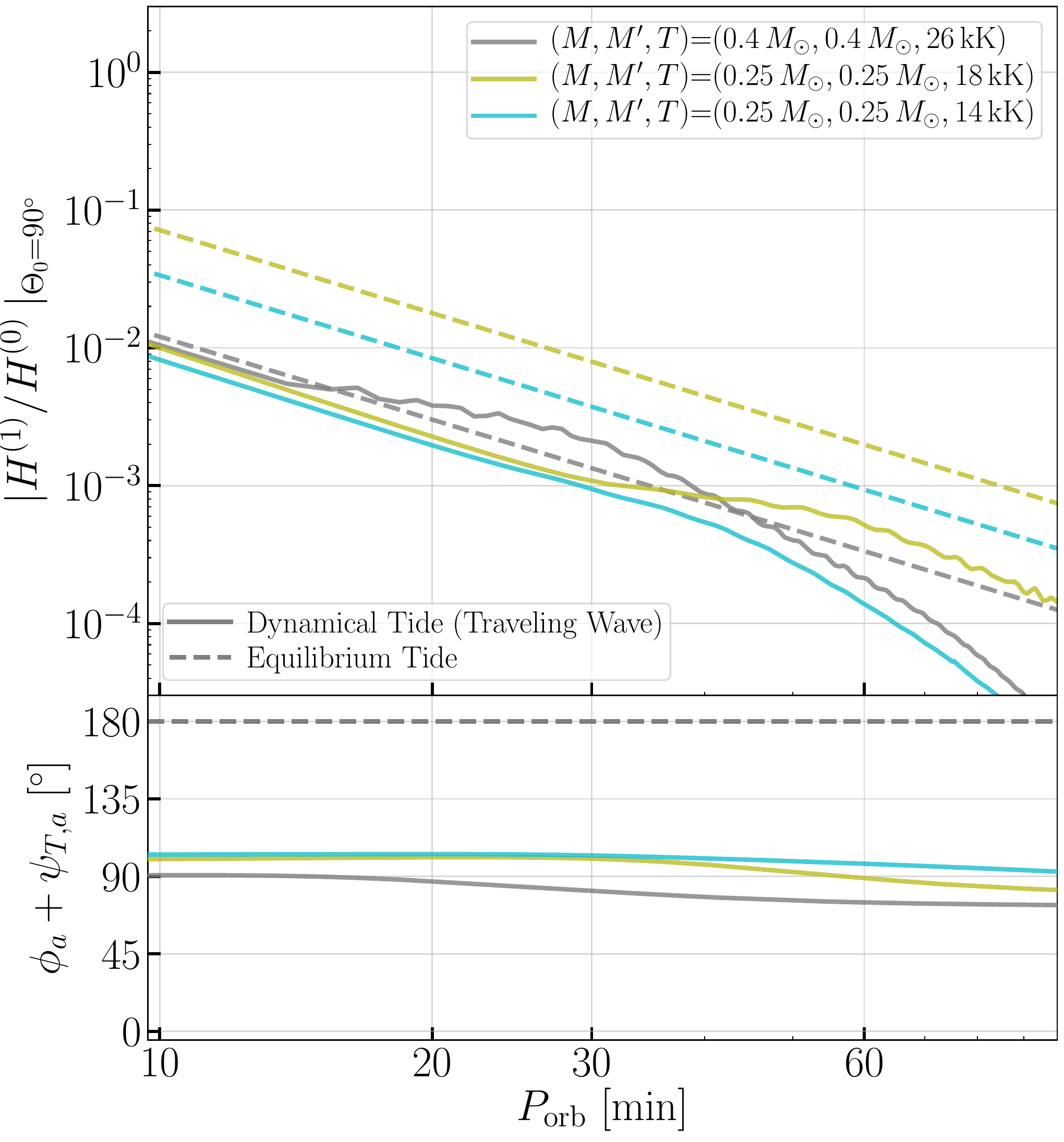} 
   \caption{Similar to Figure~\ref{fig:combo_flux_traj} but uses the traveling-wave trajectories from Figure~\ref{fig:combo_tidal_traj_trvl}}
   \label{fig:combo_flux_traj_trvl}
\end{figure}

We also consider the flux variation by treating the dynamical tide as a traveling wave (Section~\ref{sec:tw} and Figure~\ref{fig:combo_tidal_traj_trvl}) and the results are shown in Figure~\ref{fig:combo_flux_traj_trvl} (also in the dash-dotted lines in Figure~\ref{fig:M0p2_R0p06_flux_traj}). Compared to the standing-wave results, the flux variation is typically reduced by a factor of around 10-30 and is typically $0.1\%-1\%$ for $P_{\rm orb}\lesssim 20\,{\rm min}$ (a greater amplitude of $10\%$ is possible for more extended WD models as the one shown in Figure~\ref{fig:M0p2_R0p06_flux_traj}). Nevertheless, for the most compact T26 model with the smallest ellipsoidal variability, the dynamical tide can still produce a flux variation comparable to the equilibrium tide even in the traveling-wave limit and can thus be significant observationally. By construction, the excess pulsation phase is $90^\circ$ (Equation~\ref{eq:amp_a_trvl}). We point out as a caveat though that its value is largely uncertain because we do not account for local non-linearities. 

\section{Nonlinear effects}
\label{sec:nonlinearity}
As we have shown in the previous Section that the linear flux variation induced by the dynamical tide can be significant, it is thus interesting to also estimate the nonlinear effects. In particular, we study here pulsation amplitude at the $4\Omega_{\rm orb}$ harmonic, as its feature should be most prominent in the observed light curves. 

Formally, there are two nonlinear effects whose pulsation amplitudes both scale as $q_a^2$, with $q_a$ the linear mode amplitude. The first one originates from the radiative transfer process. As the temperature changes, one can expand the emerging flux $H$ [Equation~(\ref{eq:H_lambda})] as a power series of $\Delta T$. In addition to the $\left(\partial H/\partial T\right)\Delta T$ term we have used to compute the linear variation, there will also be a $\left(\partial^2 H/\partial T^2\right)(\Delta T)^2$ correction. Meanwhile, there will be nonlinear corrections to $\Delta T$ itself. In this section, we will study both effects. 

\begin{figure}
   \centering
   \includegraphics[width=0.45\textwidth]{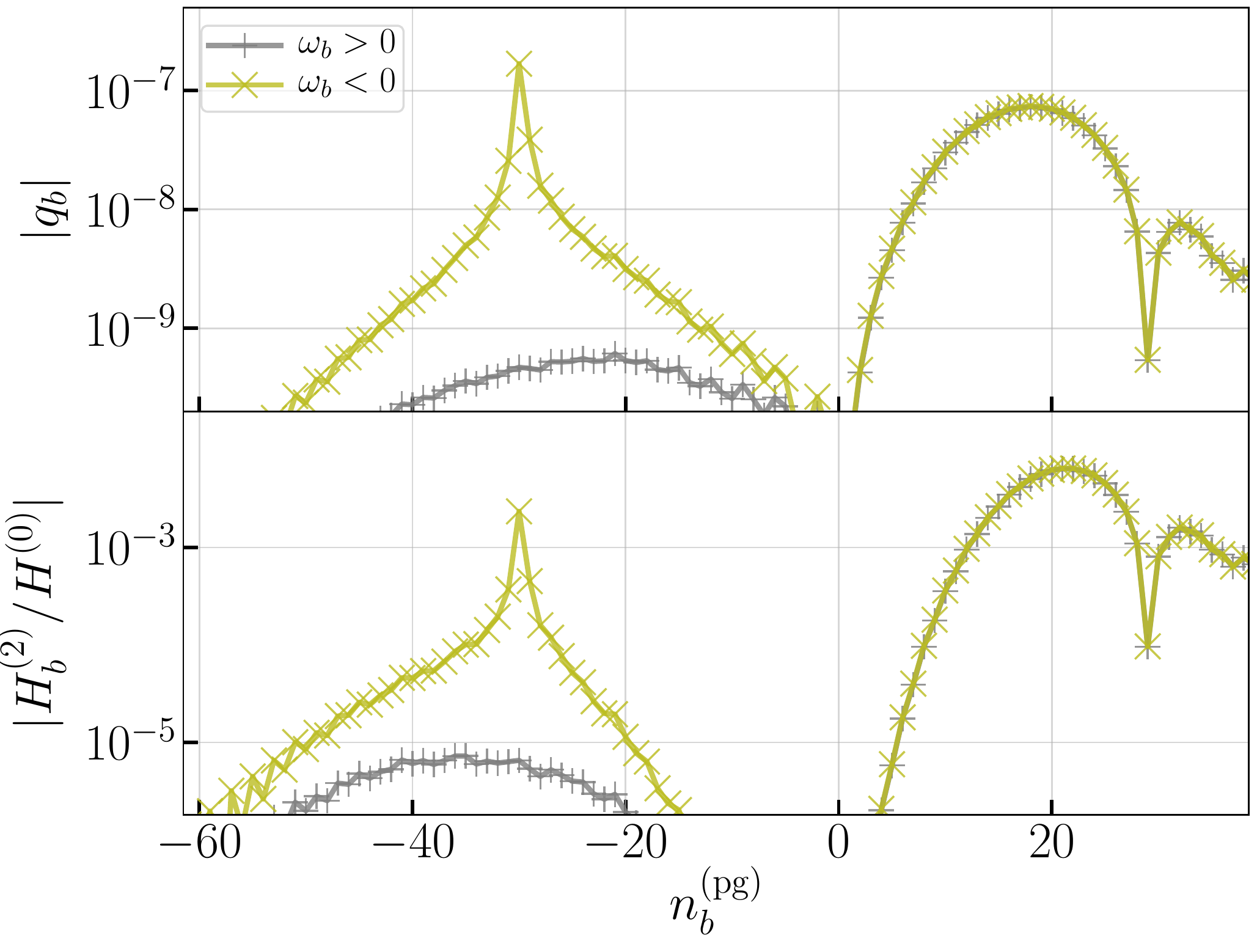} 
   \caption{Nonlinear mode amplitude (top panel) and flux (bottom panel) of daughter modes for the T26 model. The parent mode is assumed to have $n_a^{\rm (pg)}=-34$ and a mode amplitude such that $q_a \xi_a^r(R)/R=10^{-3}$. The daughters have $l_b=-m_b=4$ as required by the angular selection rule and we use grey and olive to indicate respectively modes with positive and negative frequencies.}
   \label{fig:nl_flux_vs_n_b}
\end{figure}

\begin{figure}
   \centering
   \includegraphics[width=0.45\textwidth]{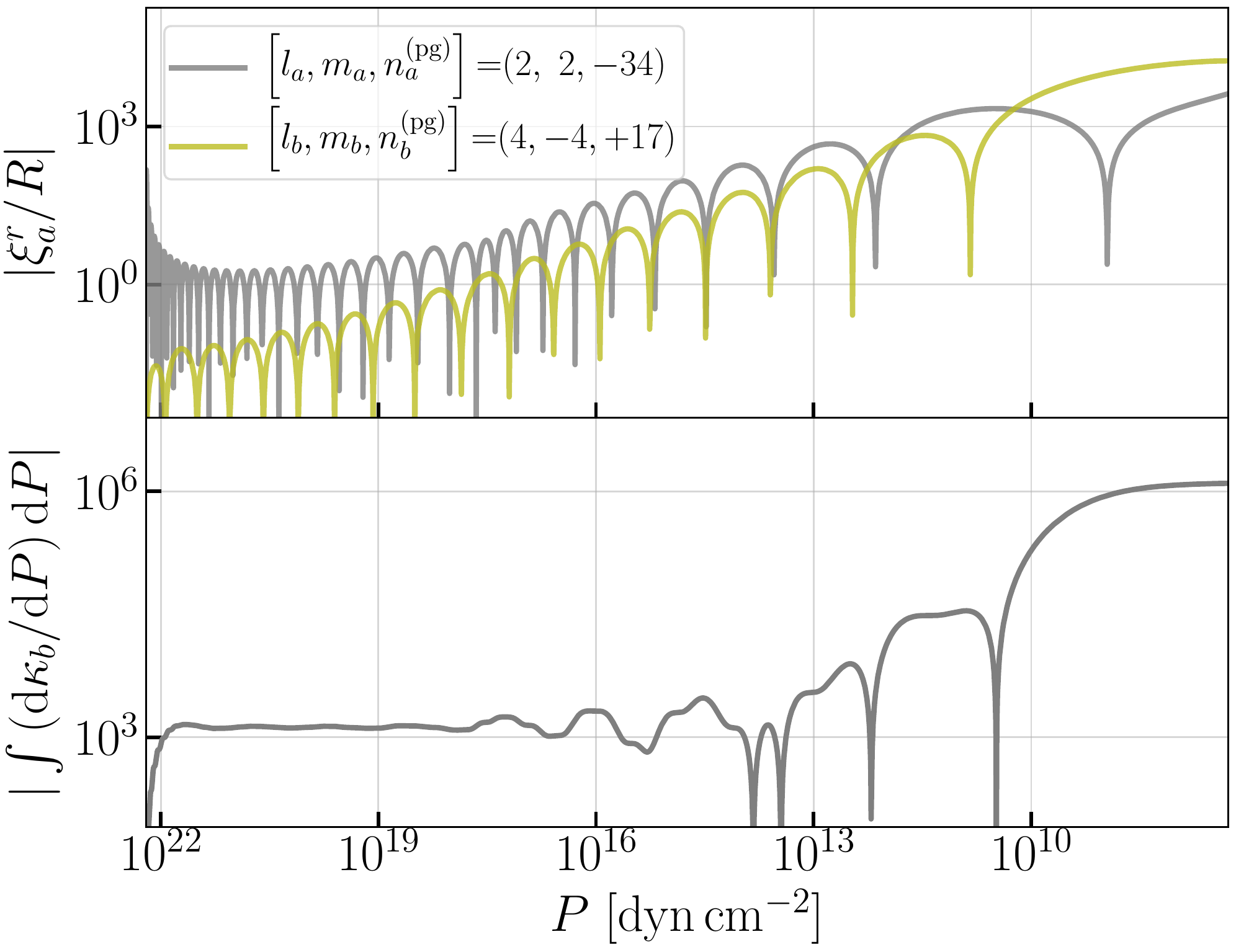} 
   \caption{Normalized radial displacements (top panel) and the cumulative three-mode-coupling coefficient (top panel). We use T26 as the background model. The parent mode is an $n_a^{\rm (pg)}=-34$ g-mode and the daughter is an $n_b^{\rm (pg)}=+17$ p-mode. Note we have inverted the x-axis so the radius increases towards right. The cumulative coupling follows the oscillation pattern of the daughter mode in the $P\leq 10^{13}\,{\rm dyn\,cm^{-2}}$ region as the self-coupled parent always has a spatially non-oscillatory component. }
   \label{fig:ggp_coup}
\end{figure}

To model the flux due to the $\left(\partial^2 H/\partial T^2\right)(\Delta T)^2$ term, we follow the prescription outlined in \citet{Brassard:95} for ZZ Ceti stars. If we assume $\left(\partial^2 H_\lambda /\partial T^2\right)/H_\lambda\simeq \left(\partial^2 I_\lambda/ \partial T^2\right)/I_\lambda$, then after some manipulation we find
\begin{equation}
    \frac{|H^{(2)}/H^{(0)}|}{|H^{(1)}/H^{(0)}|^2} = \left(\frac{|b_4|}{|b_2|^2}\right)\left(\frac{\beta^{(2)}}{64 \left[\beta^{(1)}\right]^2}\right).
\end{equation}
Here $b_2$ and $b_4$ are disc-integration factors [Equation~(\ref{eq:b_l})] evaluated at $l=2$ and $l=4$, respectively. We note that if $\Delta T_a$ has an angular pattern that corresponds to the $l=m=2$ harmonic, then $(\Delta T_a)^2$ will have $l=m=4$ for the component that varies at $4\Omega_{\rm orb}$ in the inertial frame. 

We have also defined
\begin{equation}
    \beta^{(2)}=\frac{T^2 \int  W(\lambda)  \left[\partial^2 H_\lambda^{(0)}/\partial T^2 \right] \diff\lambda}{\int W(\lambda)H_{\lambda}^{(0)} \diff \lambda},
    \label{eq:beta_2}
\end{equation}
and its value is summarized in Table~\ref{tab:beta}. Numerically, we have 
\begin{equation}
    \frac{|H^{(2)}/H^{(0)}|}{|H^{(1)}/H^{(0)}|^2} = 6.5\times10^{-3} \left[\frac{\beta^{(2)}}{0.32}\right]\left[\frac{\beta^{(1)}}{0.39}\right]^{-2}.
\end{equation}
We thus see that this effect is typically small due to both strong disc attenuation (hence small $b_4$), and optical observatories being insensitive to $\left.\partial^2 H_\lambda^{(0)}/\partial T^2 \right.$ especially for hot WDs (hence small $\beta^{(2)}$).

We estimate the nonlinear correction to $\Delta T$ itself by first computing a nonlinear displacement~\citepalias{Weinberg:12} and then use Equation~(\ref{eq:dT_from_xi}) to calculate the temperature fluctuation associated with this nonlinear displacement.\footnote{We ignore the nonlinear temperature variation induced by a linear displacement, i.e., the nonlinear correction to Equation~(\ref{eq:dT_from_xi}; see, e.g., \citealt{Brickhill:92}). Such a term is likely to be subdominant as argued in \citet{Brassard:95}. It also does not experience the strong mechanical coupling between a self-coupled g-mode and a low-order p-mode which we find plays a crucial role in making the nonlinear $\Delta T$ to be significant. 
}
Specifically, we let the  most resonant parent mode $a$ first self-couple and then inhomogeneously drive a series of daughter modes (all labeled as mode $b$). 
Note that this is not a nonlinear instability that requires the parent mode reaching a certain threshold energy. Rather, it is a nonlinear correction to the dynamics that always happens. 
Utilizing the angular selection rule and the fact that the parent mode has $l_a=m_a=2$, the daughter modes will have $l_b=-m_b=4$.\footnote{Generically, the angular selection rule requires (i) $|l_b - l_c| \leq l_a \leq l_b+l_c$, (ii) $l_a+l_b+l_c$ is even, and (iii) $m_a+m_b+m_c=0$. We note that the parent mode can also couple to itself and drive a daughter mode with $m_b=0$ and $l_b=0,\ 2,\ \text{or } 4$. Such a mode won't affect the temporal variation of the light curve, yet it may still have significant observational consequence. See the discussion at the end of this Section.} The daughter can be a g-, f-, or p-mode and its eigenfrequncy can be either positive or negative. In the co-rotating frame, the daughter mode's amplitude equation is given by~\citepalias{Weinberg:12, Yu:20} 
\begin{equation}
    \dot{q}_b + (\imag \omega_b + \gamma_b) q_b = \imag \omega_b \kappa_b q_a^\ast q_a^\ast,
\end{equation}
where $\kappa_b$ is the three mode coupling coefficient whose explicit expression can be found in \citetalias{Weinberg:12}. We have dropped the direct tidal forcing on the daughter [Equation~(\ref{eq:U_a})] because we find it is typically orders of magnitude smaller than the driving from the dynamical-tide parent.  

The daughter's steady-state solution can be obtained by noticing the asynchronicity stays nearly constant in the post-synchronization regime and $q_a\sim \exp\left(-\imag \omega_a t\right)\simeq \exp\left[-\imag m_a(\Omega_{\rm orb} - \Omega_{\rm s})\right]$ in the co-rotating frame.\footnote{Note that the detuning $(2\omega_a + \omega_b)$ is not affected by Doppler shift thanks to the condition $2m_a+m_b=0$ required by the angular section rule. The daugther mode in the inertial frame will oscillate as $\exp\left(-\imag m_b \Omega_{\rm orb}\right)$.} We thus have 
\begin{equation}
    q_b = \frac{\imag \omega_b}{\imag(2\omega_a + \omega_b) + \gamma_b}\kappa_b q_a^\ast q_a^\ast. 
\end{equation}
We further ignore the daughters' back-reaction on the parent and still use the tidal trajectories obtained under the linear approximation with $q_a$ given by Equation~(\ref{eq:mode_amp_post_sync}). We can thus compute the amplitude for each daughter $q_b$ as function of the orbital period. Similar to the linear problem, we first obtain the daughter mode's structure ($\kappa_b$, $\xi^r_b$, etc.) using the adiabatic solutions to the stellar oscillation equations and only at the end use the nonadiabatic solution to relate $\Delta T_b/T$ to $\xi_b^r(R)/R$. By summing over all the daughter modes (and their complex conjugates), we thereby compute the nonlinear correction of $\Delta T$. 

In Figure~\ref{fig:nl_flux_vs_n_b} we show the amplitude and fractional flux variation of daughter modes as a function of their radial order $n_b^{\rm (pg)}$ for the hottest T26 model. The parent mode is assumed to have a radial order $n_a = 34$ and a surface displacement $|q_a\xi_a^r(R)/R|=10^{-3}$. 
In order to distinguish a p-mode daughter from a g-mode one, we label a g-mode with a negative radial order $n_b^{\rm (pg)}<0$ while a p-mode with $n_b^{\rm (pg)}>0$; an f-mode corresponds to $n_b^{(\rm pg)}=0$~\citep{Takata:06}. We further include a superscript ``(pg)'' so that this labeling scheme should not be confused with the one used elsewhere where we consider g-modes only and a positive number is used to label to radial order. We further use the color grey (olive) to denote daughter modes with $\omega_b > 0$ ($\omega_b < 0$). 

In the g-mode branch, we note that both the mode amplitude and the flux variation is dominated by a single g-mode with $|n_b^{\rm (pg)}|\simeq n_a$ and $\omega_b<0$. This is because for a high-order g-modes, its frequency follows $|\omega_a| \sim l_a / n_a$, and consequently an $l_b=2l_a$ daughter with $n_b = |n_b^{(pg)}|\simeq n_a$ typically has a small frequency detuning with respect to the parent's inhomogeneous driving. However, the resonance is not exact and we find typically $|\omega_b/(2\omega_a+\omega_b)|\lesssim 1000$. 

On the other hand, we find a nearly equal contribution from positive- and negative-frequency modes in the p-mode brunch (with $n_b^{\rm (pg)}>0$) due to the fact that $|\omega_b|\gg \omega_a$ for p-mode daughters. Interestingly, even without temporal resonance, we still find multiple p-modes with $n_b^{\rm (pg)}\sim 20$ are driven to significant amplitudes comparable to the amplitude of the most resonant g-mode daughter. This is because of their strong spatial overlap to the parent mode which we demonstrate in Figure~\ref{fig:ggp_coup}. For a hot WD model like T26 we consider here, the parent g-mode can extend sufficiently close to the surface where the p-mode oscillation is significant. While a single parent g-mode is spatially incoherent with the daughter, as it self-couples there will always be a spatially constant component.\footnote{Conceptually, this is similar to the situation studied in \citet{Weinberg:13}, yet there are also critical differences. In our case, what is spatially coherent is a parent mode with itself instead of a pair of daughters. The process we consider is not a non-linear instability, but rather, a non-linear correction to the tidal problem that always happens without requiring any threshold. } The daughter p-mode overlaps with this constant component and thus leads to a significant accumulation of the three-mode-coupling coefficient $\kappa_b$ in its last half oscillation cycle where its displacement is the greatest. In fact, we find the total nonlinear flux is dominated by the non-resonant p-mode daughters for hot WD models.

\begin{figure}
   \centering
   \includegraphics[width=0.45\textwidth]{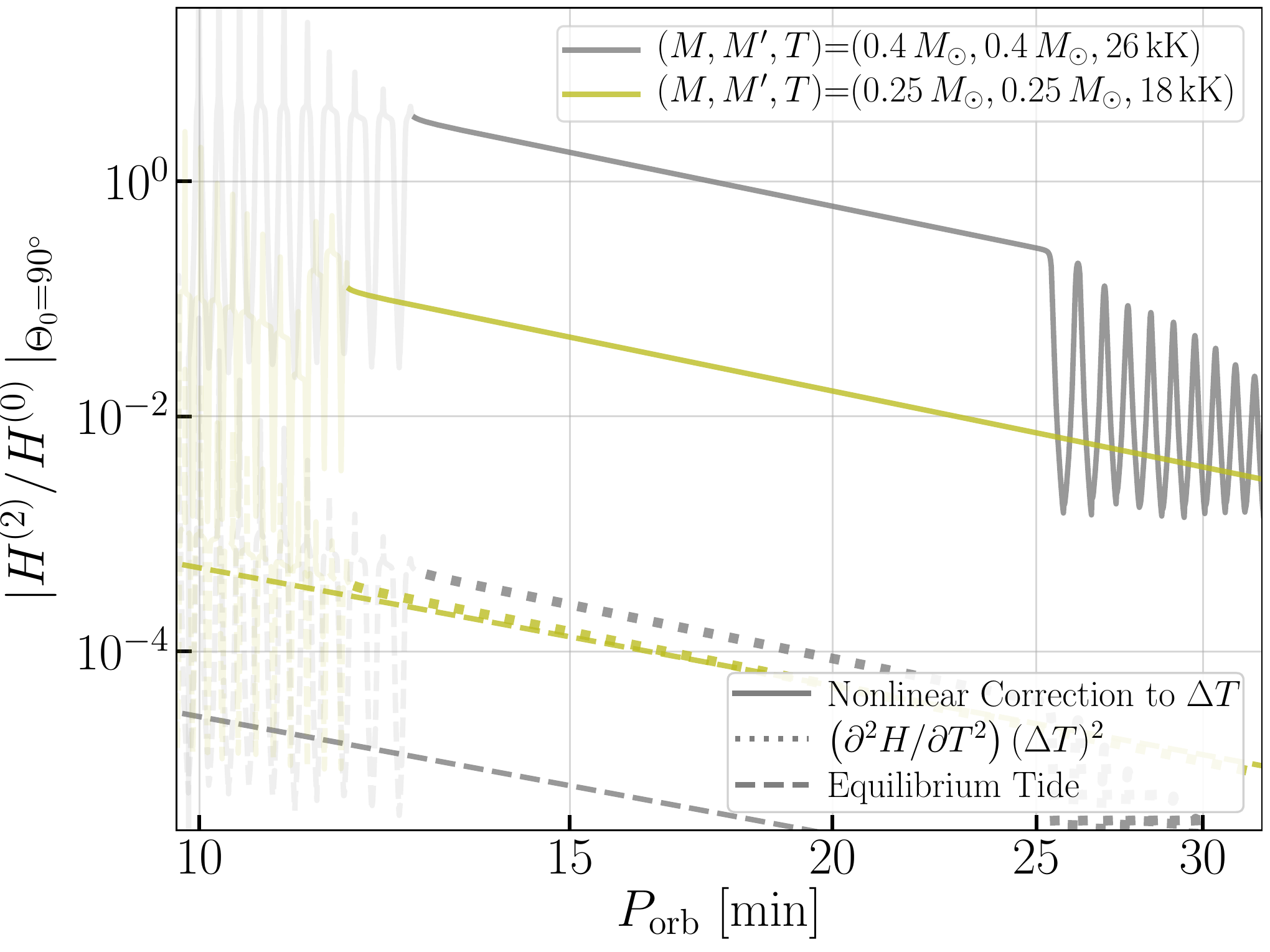} 
   \caption{Nonlinear flux variation based on the linear tidal evolution trajectory shown in Figure~\ref{fig:combo_tidal_traj}. Here we focus on the component that varies at $4\Omega_{\rm orb}$ in the inertial frame specifically.}
   \label{fig:nl_flux_traj}
\end{figure}

In Figure~\ref{fig:nl_flux_traj} we present the amplitude of the non-linear flux evolution based on the parent modes' linear tidal trajectories shown in Figure~\ref{fig:combo_tidal_traj}. We use solid, dotted, and dashed traces to represent the pulsation amplitude due to, respectively, the non-linear corrections to $\Delta T$ (particularly, $\Delta T$ from non-linear displacements), the non-linear radiative transfer process [the $(\left(\partial^2 H/\partial T^2\right)(\Delta T)^2)$ term], and the linear $l=m=4$ ellipsoidal variability (Equation~\ref{eq:ellip_var}; note we have also included its complex conjugate to physical amplitude), with the first one significantly dominating the total non-linear flux. 

More importantly, we find the nonlinear flux can in fact reach a significant amplitude. For the hottest T26 model (grey traces), the nonlinear flux actually exceeds the linear one. Even the T18 model has a nonlinear flux that is $10\%$ the linear one at $P_{\rm orb}=20\,{\rm min}$ and its significance increases as the orbit decays to shorter orbital period.

As a caveat, our results in Figure~\ref{fig:nl_flux_traj} are likely overestimating the non-linear flux, as we may already overestimating the linear flux (hence the parent mode's amplitude) by a factor of a few due to the various simplifications we adopted in the analysis. It is also uncertain when the standing-wave treatment of the parent mode ceases to be valid. Nevertheless, Figure~\ref{fig:nl_flux_traj} suggests the non-linear effects can in principle be significant and deserve further investigation both theoretically and observationally. For example, as indicated in Figure~\ref{fig:nl_flux_vs_n_b}, the combined energy in the daughter modes can reach $\sum_b |q_b|^2\simeq \text{a few}\times 10^{-14}$, which is more than $10\%$ of the parent mode energy of $|q_a|^2\sim 10^{-13}$. Therefore, their back-reaction on the parent mode may thus be significant and should be properly incorporated into the theoretical tidal evolution calculations. On the observational side, Figure~\ref{fig:nl_flux_traj} suggests that for a hot WD in a binary with orbital period around 20 minutes, the light curve may significantly deviate from a sinusoid as predicted by the linear theory due to the $4\Omega_{\rm orb}$ component. The detection (or null-detection) of such a feature could then in turn greatly constrain our understanding of tidal interactions in WDs. 

While the direct estimation of the nonlinear flux is largely uncertain, we nonetheless expect the ratio
\begin{equation}
    \frac{|H^{(2)}/H^{(0)}|}{|H^{(1)}/H^{(0)}|^2} \simeq     \begin{cases}
    47,\quad\text{ for T26}\\
    3,\quad \text{ for T18},
    \end{cases}
    \label{eq:H2_H1sq_ratio_ggp}
\end{equation}
to be a relatively robust prediction.
Note that in the expression above the linear mode amplitude $q_a$ cancels out, so does the viewing angle. In fact, the ratio depends only on the asynchronicity (determining the parent mode) and the internal WD structure (values of $\kappa_b$, $f_{T,b}$ of low-order p-modes, etc.). Unlike the mode amplitude, the asynchronicity changes only by less than a factor of 2 from the standing-wave case to the traveling-wave one (Figures~\ref{fig:combo_tidal_traj} and \ref{fig:combo_tidal_traj_trvl}). Moreover, from the lower panel of Figure~\ref{fig:ggp_coup} we note the three-mode coupling coefficient $\kappa_b$ has accumulated most of its value before the last node (counting from the center towards the surface) of parent mode. Consequently, even if the parent mode breaks at smaller pressure into traveling waves, the value of $\kappa_b$ should not be significantly altered. As a result, Equation~(\ref{eq:H2_H1sq_ratio_ggp}) should hold for both a standing-wave parent and traveling-wave one.

Equation~\ref{eq:H2_H1sq_ratio_ggp} suggests that for the T26 model, we have
\begin{equation}
    \frac{|H^{(2)}|}{|H^{(1)}|} 
    \simeq  0.47 \left(\frac{|H^{(1)}/H^{(0)}|}{0.01}\right). 
    \label{eq:H2_H1_ratio}
\end{equation}
Therefore, even if the parent's amplitude is attenuated as it becomes a traveling wave, the nonlinear flux at $4\Omega_{\rm orb}$ can still be a substantial fraction (almost half) of the linear flux. We would thus predict the light curves to exhibit significant non-linear distortions in general for very hot WDs.

So far our discussions have been focused on the component that oscillates at $4\Omega_{\rm orb}$ (with $|m_b|=4$) and it is mostly driven by a parent mode self-coupling. Similarly, a parent mode could couple with its complex conjugate and drive $m_b=0$ daughters to similar amplitudes. This coupling is a ``DC'' effect and does not affect the temporal variability of a light curve. It nevertheless has significant observational consequences because the coefficients $\beta^{(1)}$ and $\beta^{(2)}$ both depends on the bandpass of a filter, $W(\lambda)$. When $T=26\,{\rm kK}$, we find $\beta^{(1)}$ is $20\%$ greater for a filter sensitive to the $450\,{\rm nm}<\lambda<550\,{\rm nm}$ band (``g-band'') than a filter sensitive to the $725\,{\rm nm}<\lambda<950\,{\rm nm}$ band (``i-band''). Furthermore, for low-order p-modes that dominates the nonlinear flux, we find the temperature typically has an opposite sign of the displacement. As a result, the nonlinear corrections at DC would reduce the flux in the short-wavelength g-band more than in the long-wavelength i-band, making the WD appear cooler than the one without dynamical tides. In fact, this process can be intuitively understood as the following. As the dynamical tide (the parent g-mode) reaches a sufficiently large amplitudes, it causes an excess, nonlinear expansion of the WD (corresponding to the rising of daughter p-modes' amplitudes at DC). As the WD expands, its effective temperature cools accordingly. 

Since the flux from the $|m_b|=4$ component might reach a large fraction of the unperturbed flux, that from the $m_b = 0$ component is likely to also be considerable (or even more as $b_0>b_4$). Consequently, we may expect the observed temperature of a WD to be significantly modified by the nonlinear dynamical tide. A quantitative study of this effect is deferred to future work that would need to self-consistently incorporate the nonlinear corrections to the tidal evolution first. 



\section{Discussion}
\label{sec:discussion}

As we have predicted that the dynamical tide could lead to significant perturbation to the observed flux, it would be nice to compare them with existing observations. We note that there are two 20-min He WD binaries discovered recently, J0533~\citep{Burdge:19b} and J2322~\citep{Brown:20}, that are potentially interesting. 

In J0533, the visible object ($M_{\rm B}$ in the notation of \citealt{Burdge:19b}) has a temperature similar to our T18 model but is more extended with a greater radius of $0.06\,R_\odot$. The authors reported a significant ellipsoidal variability at $\mathcal{O}(10\%)$ level and the pulsation phase matches nicely the expectation of the equilibrium tide~\citep{Burdge:19b}. If there were a resonant mode (i.e., a standing wave) in the visible object, we would expect it to produce a similar (or even greater) level of photometric modulation with a shifted phase (see Figure~\ref{fig:combo_flux_traj}). The null detection of this feature thus rules out this possibility and favors the traveling-wave picture.  Indeed, this is expected because the visible object  has a large radius of $0.06\,R_{\odot}$. As we show in Figure~\ref{fig:M0p2_R0p06_flux_traj} (whose parameters are similar to the visible object in J0533), for such an extended WD the equilibrium tide is expected to be significant at the $\mathcal{O}(10\%)$ level, whereas the dynamical tide is likely in the wave-breaking regime and hence subdominant. 

The primary in J2322 (the DA WD) is more similar to our T18 model. However, this system is viewed nearly pole-on and therefore does not show a prominent pulsation signature at  $2\Omega_{\rm orb}$ but an upper limit of $\lesssim 0.3\%$ is reported in \citet{Brown:20}. If we use the most-likely inclination angle of $\Theta_0=27^\circ$ as reported in \citet{Brown:20}, then our T18 model would predict a flux of $1\%$ due to the dynamical tide. If the upper limit is near the true level of the dynamical tide, then this would suggest that we have overestimated the result by a factor of 2-3, potentially due to our neglecting of the rotational correction \citep{Fuller:14} and/or the nonlinear effects (especially the inhomogeneous coupling to the daughter modes that always happens; see Section~\ref{sec:nonlinearity}). Alternatively, 
if the upper limit on the dynamical tide's flux can be further lowered by an order of magnitude with future observations (especially with the phase resolved to distinguish it from and ellipsoidal variability), then it would suggest the tidal wave breaks at much lower threshold than Equation~(\ref{eq:wave_break_cond}).


\begin{figure}
   \centering
   \includegraphics[width=0.45\textwidth]{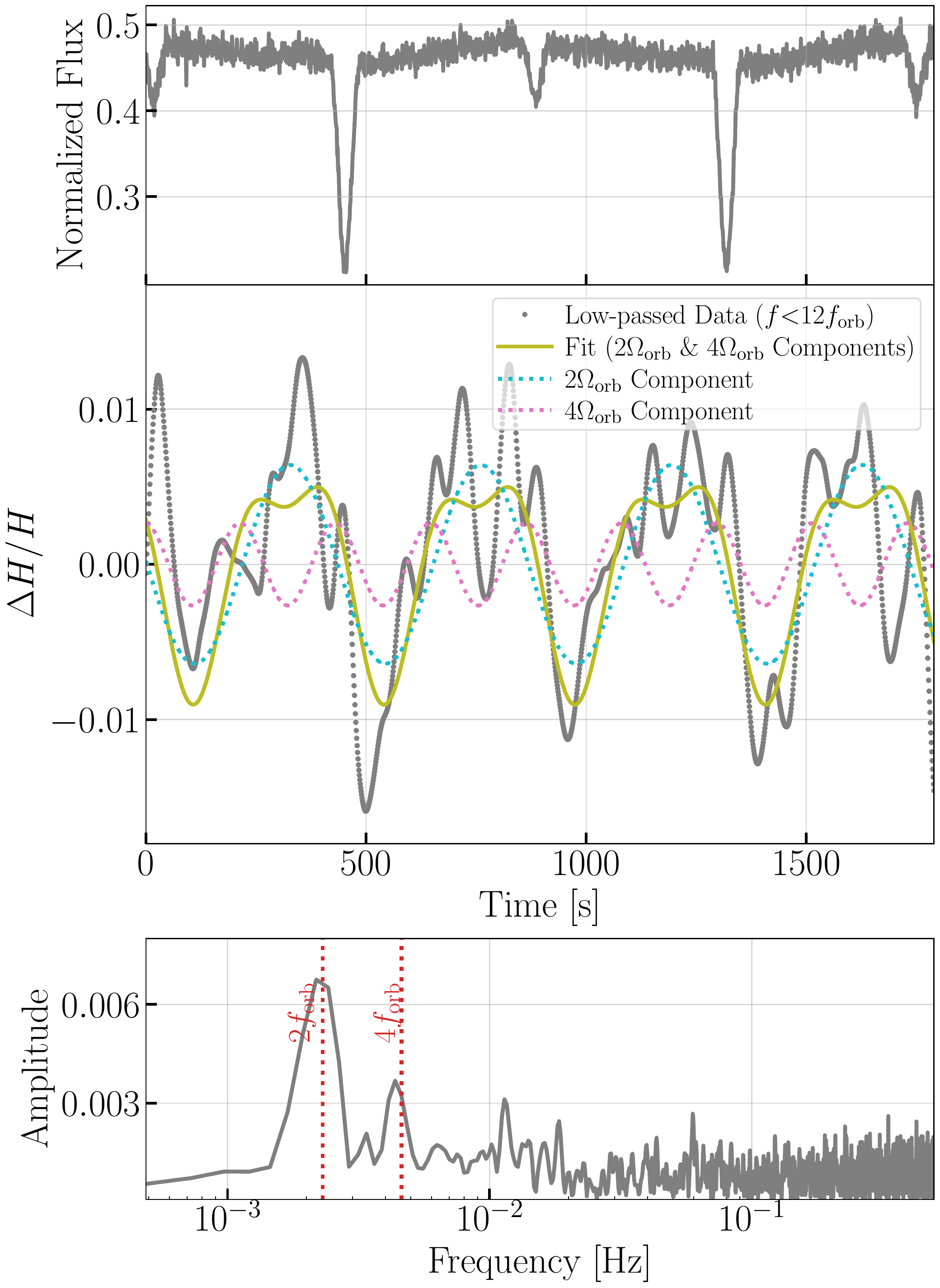} 
   \caption{Top: full light curve of the 14-min WD binary J0538~\citep{Burdge:20}, collected using HiPERCAM on the 10.4
meter Gran Telescopio Canarias~\citep{Dhillon:18}.
   Middle: fractional residual after the removal of the eclipsing signals (grey dots). For visualization, we have low-passed the data to keep only the low-frequency components. The olive trace is the time-domain, least-square fit of the variation (using the full residual), including a $2\Omega_{\rm orb}$ component and a $4\Omega_{\rm orb}$ one. Each individual component is shown in the dotted trace. The starting time corresponds to Barycentric Julian Date in the Barycentric Dynamical Time of 58734.24. 
   Bottom: amplitude spectrum of the residual. 
   The peak corresponds to the amplitude (instead of spectral density) of each frequency harmonic. 
   }
   \label{fig:J0538}
\end{figure}

A more exciting system is the recently reported 14-min binary J0538~\citep{Burdge:20}, whose hotter component ($M_{\rm A}$) is similar to our T26 model. In the top panel of Figure~\ref{fig:J0538}, we show the full light curve of the system. The fractional residual after removing the eclipsing signals is shown in grey dots in the middle panel. Note we have low-passed the data to show only components with frequency $f<12f_{\rm orb}$, where $f_{\rm orb}=\Omega_{\rm orb}/2\pi$. Clearly, we see the data points are consistent with a sinusoidal variation at $2\Omega_{\rm orb}$ distorted by higher-frequency harmonics. We thus fit the residual data with two sine waves oscillating respectively at $2\Omega_{\rm orb}$ and $4\Omega_{\rm orb}$ and the least-square fitting result is shown in the olive-solid trace. We also show the individual frequency component respectively in the cyan-dotted and pink-dotted traces. In the bottom panel, we present the amplitude spectrum (not spectral density) of the fractional residual. We also mark the locations of $2f_{\rm orb}$ and $4f_{\rm orb}$ with vertical-dotted lines. 

One of the most critical features to note is the phase of pulsation. Note that the $2\Omega_{\rm orb}$ component crosses 0 when the binary eclipses, suggesting an excess pulsation phase of $90^\circ$. This is in contrast to what we would expect for modulations due to the equilibrium tide, which should have troughs at the eclipses as one can show from a simple geometrical argument. Such a phase shift is the reason why \citet{Burdge:20} found unphysical mass ratios if they tried to fit the flux with ellipsoidal variation. It thus strongly hints at a dynamical-tide origin of the pulsation. 

On the other hand, both the time-domain least-square fit and the frequency-domain spectral analysis suggest a fractional variation of $|H^{(1)}/H^{(0)}|=0.6-0.7\%$ for the $2\Omega_{\rm orb}$ component. This is about a factor of 30 smaller than our prediction shown in Figure~\ref{fig:combo_flux_traj} assuming the standing-wave picture applies. It thus suggests that the wave in fact breaks already at $P_{\rm orb}\simeq 14\,{\rm min}$, a period similar to but slightly greater than our estimate (see Figure~\ref{fig:combo_flux_traj}).
It further suggests that Equation~(\ref{eq:wave_break_cond}) potentially overestimates the amplitude required to enter this strongly nonlinear regime (see also the discussions in \citealt{Fuller:13}). 

If we instead use the prediction based on the traveling-wave limit, then we would predict a flux variation of $0.6\%$ according to Figure~\ref{fig:combo_flux_traj}. Moreover, it predicts a pulsation phase of around $90^\circ$ relative to the orbit (Equation~\ref{eq:amp_a_trvl}). While both the amplitude and the phase seem to match the observation, we note the traveling-wave interpretation should still be taken with caution. For example, we have ignored all the highly nonlinear processes and still use the linear $f_{\rm T, a}$ and $\psi_{\rm T, a}$ to relate the temperature variation to the displacement (Equation~\ref{eq:dT_from_xi}). The validity of this approximation should be tested with future observations. 

Beyond the $2\Omega_{\rm orb}$ component $H^{(1)}$, we also find a significant $4\Omega_{\rm orb}$ component $H^{(2)}$ that creates about $0.3\%$ fractional modulation of the flux. Since the standing-wave estimate of the parent mode's flux already exceeds the observation by a factor of 30, we do not expect Figure~\ref{fig:nl_flux_traj} to hold for J0538. Nevertheless, we find the ratio
\begin{equation*}
    \frac{|H^{(2)}/H^{(0)}|}{|H^{(1)}/H^{(0)}|^2} \simeq 70-80,\quad \left(\text{or } \frac{|H^{(2)}|}{|H^{(1)}|} \simeq 0.4\right),
\end{equation*}
from the observation, which matches our prediction on the nonlinear flux due to the parent's inhomogeneous coupling to daughter modes (mostly low-order p-modes) within a factor of 2 (see Equations~\ref{eq:H2_H1sq_ratio_ggp} and \ref{eq:H2_H1_ratio}). As we argued in Section~\ref{sec:nonlinearity}, this ratio should be a relatively robust prediction because it is independent of the parent mode's amplitude to the leading order and should be insensitive to whether the parent is a standing wave or a traveling wave. It can thus be used to test the mechanism we proposed observationally. The agreement shown by J0538 is promising, and we look forward to further validating it with future observations. 


Looking towards future, it would be of great value to characterize more light curves of compact WD binaries observationally. For WDs in binaries with $20\,{\rm min} \lesssim P_{\rm orb}\lesssim 60\,{\rm min}$, the linear, standing-wave approximation we adopted may apply if the temperature $T\gtrsim 14\,{\rm kK}$. Either a clean detection or a definitive null detection of the dynamical tide signature would greatly improve our understanding of the tidal evolution history of these systems. The more compact binaries with $P_{\rm orb}\lesssim 20\,{\rm min}$ are more likely to excite tidal waves as highly nonlinear traveling waves. While our crude approximation of tides in this regime appears to match the observed properties of J0538, the validity of this approximation remains to be tested with more observations. The nonlinear flux variation at $4\Omega_{\rm orb}$ is another key feature to look for. Of particular interest is the ratio between this nonlinear flux modulation and the square of the linear one, which distinguishes the mechanism producing the nonlinear effect. 

On the theoretical side, one of the greatest uncertainties is the nonlinear effect. For the weakly nonlinear systems in which the parents can be still treated as standing waves, even without nonlinear instabilities, the inhomogenous coupling to daughter modes may already be important. For example, the daughter modes can also contribute to the dissipation of energy and thus increase the effective damping rate as the energy in the parent mode increases. If a resonant lock to a particular parent mode is maintained, then Equation~(\ref{eq:mode_amp_post_sync}) suggests the parent mode's amplitude will be reduced relative to the linear case due to the increased $\gamma_a(P_{\rm c})$, thereby further reducing the flux variation. If this energy-dependent nonlinear damping dominates over the constant linear damping, then the resonant lock will break, similar to the case shown in \citetalias{Yu:20}. 

The exact transition location from the weakly  nonlinear to the strongly nonlinear regimes is also of great significance. The observations of J0538~\citep{Burdge:20} (and potentially J2322;~\citealt{Brown:20}) seem to suggest that gravity waves break before Equation~(\ref{eq:wave_break_cond}). We note that the wave-breaking condition has been studied extensively in the context of solar-type stars where the maximum shear happens at the core~(e.g., \citealt{Ogilvie:07, Barker:11}). Recently, \cite{Su:20} also consider this in the context of WD envelopes using a panel-parallel geometry. All the studies suggest a wave-breaking condition consistent with Equation~(\ref{eq:wave_break_cond}). Nonetheless, as the tidal excited gravity wave deposits angular momentum, it leads to the formation of a critical layer separating out an upper, synchronized region near the surface from the lower, stationary region~\citep{Su:20}. This layer may further propagate from the surface downwards, which potentially makes the waves increasingly harder to perturb the surface temperature and consequently the emergent flux. 
Furthermore, \cite{Su:20} suggests that such a layer might be partially reflecting and consequently modify the dynamics in the strongly non-linear regime as well as the tidal flux perturbations. More detailed examinations of its consequences are warranted. 

\section{Conclusion}
\label{sec:conclusion}

We studied dynamical tides in short-period WD binaries, examining the flux perturbations due to tidally excited gravity waves in WDs with various surface temperatures.

Because gravity mode damping rates increase with the WD temperature, the maximum amplitude reached at resonance is smaller for hotter WD models. Consequently, the spin becomes synchronized with the orbit at shorter orbital periods (that is, later in the orbital evolution) for hotter WD models. For WDs with $14\,{\rm kK}<T<26\,{\rm kK}$, we find the synchronization happens at an orbital period $P_{\rm c}\simeq 25-60\,{\rm min}$. Once synchronization begins, the resonantly locked mode maintains a constant asynchronicity, until the mode dissipates or breaks due to nonlinear effects. 

The dynamical tide can typically produce more than a $1\%$ modulation of the observed flux in this standing-wave regime for hot WD models with $T\gtrsim 14\,{\rm kK}$, where gravity waves can propagate sufficiently close to the surface. The flux variation due to the dynamical tide can be more significant than the ellipsoidal variability if the WD is compact (with $R\lesssim 0.04 R_\odot$), and their ratio is independent of the viewing angle and evolves only mildly with orbital period. Furthermore, the dynamical tide's pulsation may be phase shifted by $\simeq 50^\circ$, which can be used to distinguish it from the equilibrium tide. 

As the system evolves further, the tidally excited gravity modes likely reach a large enough amplitude to break due to local nonlinear effects. We then model it as a one-way traveling wave in this strongly nonlinear regime. The orbital period for this transition is expected to be around $P_{\rm orb}\simeq 13\,{\rm min}$ for the WD model with $T=26\,{\rm kK}$, and around $P_{\rm orb}\simeq 28\,{\rm min}$ for $T=14\,{\rm kK}$, though some uncertainties remain. Once nonlinear wave-breaking occurs, the pulsation amplitude is typically reduced by a factor of 30 compared to the linear standing wave. Nonetheless, for sufficiently compact WDs with $R\lesssim 0.02\,R_\odot$, the dynamical tide's perturbation on the flux could still exceed the ellipsoidal variability, and observationally one may use the pulsation phase to differentiate the two. For example, the variations in the residual light curve of the 14-min binary J0538 (likely dominated by the hotter component $M_{\rm A}$ whose property is similar to our T26 model; \citealt{Burdge:20}) may be potentially explained by dynamical tides in the traveling-wave regime. 

We also estimated the nonlinear flux modulation that varies at $4\Omega_{\rm orb}$, which is also readily detectable in hot WDs with $T\gtrsim 26\,{\rm kK}$. It could be as large as $50\%$ of the linear one at $2\Omega_{\rm orb}$ when the fractional linear amplitude is around $1\%$. While the absolute value of the nonlinear variation is largely uncertain due to the uncertainties in the linear one already, we nonetheless find the ratio of the  nonlinear flux variation to the square of the linear variation to be a robust prediction and can be used to test the mechanisms producing the nonlinear flux variations. 

\section*{Acknowledgements}
HY acknowledges support from the Sherman Fairchild Foundation. JF is thankful for support through an Innovator Grant from The Rose Hills Foundation, and the Sloan Foundation through grant FG-2018-10515. KBB thanks the National Aeronautics and Space Administration and the Heising Simons Foundation for supporting his research.

\section*{Data availability}
The \texttt{MESA} and \texttt{GYRE} input files to generate the background models and to obtain the WD eigenmodes are available on reasonable requests to the corresponding author.

\bibliographystyle{mnras}
\bibliography{ref,RefJim}{}


\bsp	
\label{lastpage}
\end{document}